\documentclass[aps,pre, superscriptaddress, reprint, amsmath, amssymb]{revtex4-2}
\usepackage{graphicx} % Required for inserting images
\usepackage[dvipsnames]{xcolor}
\usepackage{amsmath,amssymb}
\usepackage{esint}
\usepackage{physics}
\usepackage[colorlinks]{hyperref}
\hypersetup{
	linkcolor= NavyBlue,
	citecolor=	ForestGreen,
	linktocpage=true}
\graphicspath{ {./Figs/} }

\begin{document}
\newcommand{\nuuddd}{\nu_{\uparrow\downarrow,\downarrow\downarrow}}
\newcommand{\nuuudd}{\nu_{\uparrow\uparrow,\downarrow\downarrow}}
\newcommand{\nuuddu}{\nu_{\uparrow\downarrow,\downarrow\uparrow}}
\newcommand{\nuuuud}{\nu_{\uparrow\uparrow,\uparrow\downarrow}}

\title{Thermometry by correlated dephasing of impurities in a 1D Fermi gas}
\author{Sindre Brattegard} 
\email{brattegs@tcd.ie}
\affiliation{School of Physics, Trinity College Dublin, College Green, Dublin 2, Ireland}
\author{Mark T. Mitchison}
\email{mark.mitchison@tcd.ie}
\affiliation{School of Physics, Trinity College Dublin, College Green, Dublin 2, Ireland}
\affiliation{Trinity Quantum Alliance, Unit 16, Trinity Technology and Enterprise Centre, Pearse Street, Dublin 2, D02YN67}

\begin{abstract}
We theoretically investigate the pure dephasing dynamics of two static impurity qubits embedded within a common environment of ultracold fermionic atoms, which are confined to one spatial dimension. Our goal is to understand how bath-mediated interactions between impurities affect their performance as nonequilibrium quantum thermometers. By solving the dynamics exactly using a functional determinant approach, we show that the impurities become correlated via retarded interactions of the Ruderman–Kittel–Kasuya–Yosida type. Moreover, we demonstrate that these correlations can provide a metrological advantage, enhancing the sensitivity of the two-qubit thermometer beyond that of two independent impurities. This enhancement is most prominent in the limit of low temperature and weak collisional coupling between the impurities and the gas. We show that this precision advantage can be exploited using standard Ramsey interferometry, with no need to prepare correlated initial states nor to individually manipulate or measure the impurities. We also quantitatively assess the impact of ignoring these correlations when constructing a temperature estimate, finding that acceptable precision can still be achieved in some parameter regimes using a simplified model of independent impurities. Our results demonstrate the rich nonequilibrium physics of impurities dephasing in a common Fermi gas, and may help to provide better temperature estimates at ultralow temperatures. 

\end{abstract}

\maketitle

\section{Introduction}

The last decades have seen impressive developments in experimental techniques for ultracold atomic gases. It is now possible to tune the geometry and dimensionality of the system, the nature and strength of interparticle interactions, and even the exchange statistics of the constituent particles, enabling exploration of a wide range of fascinating physical phenomena~\cite{Bloch2008rmp, Guan2013,blochquantum2012,mistakidis_cold_2022}. A particularly promising development in this regard is the creation of homogeneous ultracold gases~\cite{gaunt_bose-einstein_2013, schmidutz_quantum_2014, navon2015critical, mukherjee_homogeneous_2017, hueck_two-dimensional_2018, mukherjee_spectral_2019, yan_boiling_2019}: these are especially useful for the quantum simulation of condensed-matter and high-energy physics models, for which translation invariance is crucial. However, measuring the temperature of these systems is challenging. Standard methods like time-of-flight measurements completely obliterate the system, do not give access to spatially resolved temperature information, and may suffer from a loss of precision at ultra-low temperatures~\cite{Onofrio_2016}. Other techniques such as thermometry based on density fluctuations, although being non-destructive, may also yield low precision at ultra-low temperatures~\cite{muller2010local, zhou2011universal, hartke2020doublon}.

An interesting alternative is to work with ultracold atomic mixtures comprising more than one species of atom~\cite{sowinski_one-dimensional2019,recati_coherently_2022}. If the density of one species is much lower than the others, they may be considered as impurities embedded in a fluid of majority atoms. Such impurities a exhibit range of interesting behaviour straddling the interface between open quantum systems and condensed-matter physics. For example, mobile impurities form polarons in bosonic \cite{yegovtsev_effective_2023, skou_non-equilibrium_2021} and fermionic \cite{Massignan2014, ness_observation_2020} environments, while a static impurity exhibits universal dynamics manifesting the Anderson orthogonality catastrophe~\cite{anderson_infrared_1967, goold_orthogonality_2011, knap_time-dependent_2012,cetina_ultrafast_2016,schmidt_universal_2018}. By measuring static and dynamic properties of the impurities, numerous experiments have been able to accurately infer the temperature of the host gas~\cite{Regal2005,Spiegelhalder2009,Nascimbene2010,McKay2010,Olf2015,Hohmann2016,Lous2017,adam_coherent_2022}. This opens up the tantalising prospect of exploiting the quantum mechanical behaviour of ultracold impurities for improved thermometry, in a way that is also local and minimally destructive, in principle. 

The quest to understand the fundamental capabillities and limits of quantum thermometry has inspired a substantial theoretical literature (see Ref.~\cite{mehboudi_thermometry_2019} for a review). Seminal early work established the optimum sensitivity and level structure of fully equilibrated probes~\cite{correa_individual_2015}, while numerous proposals have put forward the possibility of using nonequilibrium impurity dynamics for thermometry, especially in the context of ultracold gases~\cite{Bruderer2006njp,Sabin2014,Hangleiter2015,Johnson2016,razavian_quantum_2019,mitchison_situ_2020,oghittu_quantum-limited_2022, mitchison_taking_2022, yuan_quantum_2023}. More recently, it has been established how temperature estimation is affected by informational constraints such as limited measurement data~\cite{rubio_global_2021,mok_optimal_2021,boeyens_uninformed_2021,jorgensen_bayesian_2022,mehboudi_fundamental_2022, alves_bayesian_2022} or coarse-grained measurements~\cite{jorgensen_tight_2020,hovhannisyan_optimal_2021}. A particularly important issue of current interest is to understand how relevant physical effects such as strong system-bath correlations~\cite{correa_enhancement_2017,miller_energy-temperature_2018,Mehboudi2019,mihailescu_thermometry_2023} may affect thermometry protocols in real impurity systems.

Another important physical effect is the interaction between several probes induced by their mutual interaction with a common thermal environment. This is important because experiments typically operate in a regime with several impurities --- perhaps even hundreds or thousands --- embedded in a single copy of the gas. Naively, increasing the number of impurities is helpful to increase the signal-to-noise ratio, which scales by a factor of $\sqrt{M}$ for $M$ independent impurities. Yet independence is generally spoiled by bath-induced interactions, which arise naturally in ultracold mixtures~\cite{shen_exact_2017,santamore_fermion-mediated_2008,windt_fermionic_2023, dean_impurities_2021, li_impurity-induced_2023, klein_interaction_2005, astrakharchik_many-body_2023,ding_mediated_2022, drescher_medium-induced_2023, addis_two-qubit_2013} and have been observed in recent experiments~\cite{edri_observation_2020, baroni_mediated_2023_nat}. Previous theoretical work has shown that, in some settings, trapped impurities can be configured to suppress bath-mediated interactions~\cite{Hangleiter2015,mitchison_probing_2016}, but in general these interactions can give rise to classical and quantum correlations between impurities~\cite{longstaff_persistent_2023}. It is well known that quantum correlations can yield a metrological advantage in some scenarios~\cite{Hyllus2012,Toth2012} and thermometry is no exception~\cite{aybar_critical_2022}. Indeed, recent works have shown that bath-mediated interactions can improve temperature estimation for impurities embedded in a bosonic environment~\cite{gebbia_two-qubit_2020, planella_bath-induced_2022,brenes_multi-spin_2023}. 

In this work, we take the first steps towards understanding how bath-induced interactions affect thermometry for impurities embedded in a fermionic environment. Following previous work by one of us~\cite{mitchison_situ_2020}, we focus on dephasing probes~\cite{razavian_quantum_2019, oghittu_quantum-limited_2022, mitchison_taking_2022, yuan_quantum_2023}, where information about the temperature is imprinted on quantum coherences that can be measured experimentally by Ramsey interferometry~\cite{cetina_ultrafast_2016,cetina_decoherence_2015,skou_non-equilibrium_2021,adam_coherent_2022}. Specifically, we consider a system of two static impurities, each possessing two internal states, which are coupled to a one-dimensional, homogeneous Fermi gas. Homogeneous systems are particularly challenging for thermometry, since absorption imaging of the spatial density provides no information on temperature, while time-of-flight imaging of the momentum distribution yields diminishing sensitivity at low temperature because only a small number of atoms near the Fermi energy yield useful information. In this context, we show that correlations induced by bath-mediated interactions can yield a collective enhancement for thermometry at low temperatures. Remarkably, this advantage survives even when one is limited to local observables that are accessible via Ramsey interferometry.  

In the following, we describe our two-impurity setup in detail (Sec.~\ref{subsec_description}) and explain how to solve the quantum dynamics of the impurities exactly using a functional determinant approach~\cite{abanin_fermi-edge_2005, dambrumenil_fermi_2005,schonhammer_full_2007} (Sec.~\ref{subsec_calcdec}). We then analyse the dephasing dynamics in detail (Sec.~\ref{impurity_dynamics}), focusing especially on the effect that the impurities have on each other via the bath. Unlike the dipole-dipole interactions that typically arise in bosonic baths, itinerant fermions induce Ruderman–Kittel–Kasuya–Yosida (RKKY) interactions between localised spins~\cite{Ruderman1954,kasuya_theory_1956} with a non-trivial oscillatory spatial dependence whose period is set by the Fermi wavelength. Static RKKY-like interactions between impurities have already been shown to arise in one-dimensional atomic Fermi gases~\cite{Recati2005casimir,fuchs2007oscillating}. Interestingly, here we show that monitoring the two-impurity dynamics allows one to observe the effect of RKKY-like interactions developing in real time.

Next, we describe how temperature can be optimally inferred from measurements on the impurities (Sec.~\ref{quant_est}), and propose a Ramsey protocol that is generally sub-optimal but experimentally feasible (Sec.~\ref{subsec_ramsey}). Our approach is based on local temperature estimation theory~\cite{paris_quantum_2009,mehboudi_thermometry_2019}, where the quantum Fisher information sets the ultimate limit for precision. Using these tools, we analyse the temperature sensitivity of the two-impurity system (Sec.~\ref{thermometric_performance}), finding that bath-induced correlations enhance precision even under local measurements, while using entangled initial states yields no advantage. 

Finally, we ask under what conditions the impurities can be modelled as independent (Sec.~\ref{subsec_indep_probes}). This approximation may be useful to simplify the construction of the temperature estimator, especially when scaling to a larger number of impurities $M>2$. For two impurities at fixed positions, we find that the validity of this approximation depends quite strongly on the impurity separation, but works best at higher temperatures. Apart from their potential relevance for current experiments on multiple impurities, our results open up several interesting future directions for theoretical research that we discuss in Sec.~\ref{subsec_indep_probes}.

\section{Setup}
\label{sec_setup}

\subsection{Description of the system}
\label{subsec_description}

\begin{figure}
    \centering
    \includegraphics[width=0.49\textwidth]{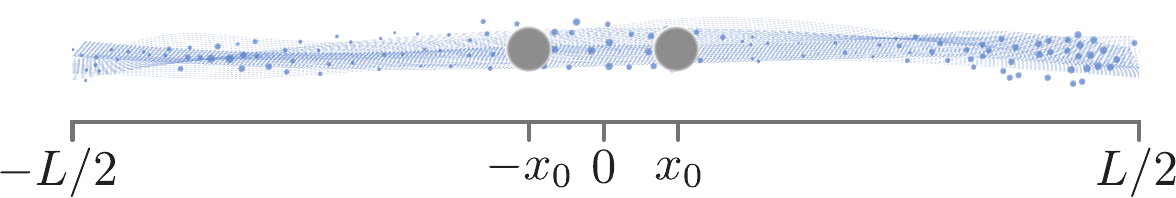}
    \caption{A sketch of the setup we are considering. Two impurities separated by distance $2x_0$ (gray balls) embedded in a 1D Fermi gas confined by a box potential of length $L$ (blue background).}
    \label{setup}
\end{figure}
We consider a system $S$ of two impurity atoms coupled to an environment $E$ comprising a gas of spin-polarized fermions with homogeneous density, which is confined to a one-dimensional (1D) box of length $L$. We are interested in the ultra-low temperature regime where $s$-wave scattering is dominant. Because of the anti-symmetry of the fermionic wavefunction there is no $s$-wave scattering between identical fermions, so the environment can be treated as a non-interacting gas to a good approximation.

The impurity atoms are modelled as two-level systems with energy eigenstates $\ket{\uparrow}_i$, $\ket{\downarrow}_i$, with $i=1,2$ labelling the two impurity qubits. We take the impurities to be stationary and strongly localized at fixed positions $x_1$ and $x_2$, which can be achieved using a species-selective optical lattice that only affects the impurities~\cite{catani_quantum_2012,adam_coherent_2022,schmidt_quantum_2018}. We work in the pseudopotential approximation, which models $s$-wave collisions between the impurities and the surrounding gas atoms as a contact interaction. The total Hamiltonian is then
\begin{align}
	\label{H_tot}
   &  \hat{H} = \hat{H}_S+\hat{H}_E+\hat{H}_I,\\ 
   \label{H_S}
   & \hat{H}_S = \sum_{i=1}^2 \varepsilon_i \ket{\uparrow}_i \!\bra{\uparrow}, \\
   \label{H_E}
   & \hat{H}_E  =  -\frac{\hbar^2}{2m}\int_{-L/2}^{L/2} \dd x\, \hat{\Psi}^\dagger(x) \nabla^2 \hat{\Psi}(x), \\
   \label{H_I}
   & \hat{H}_I = -\frac{\hbar^2}{ma} \sum_{i=1}^2 \ket{\uparrow}_i \!\bra{\uparrow} \otimes \hat{\Psi}^\dagger(x_i) \hat{\Psi}(x_i),
\end{align}  
where $\varepsilon_i$ is the local energy splitting between the impurity eigenstates, $\hat{\Psi}^{\dagger}(x)$ is the field operator that creates a fermion of mass $m$ at position $x$, such that $\{\hat{\Psi}(x), \hat{\Psi}^\dagger(x')\} = \delta(x-x')$, and $a$ is the scattering length describing impurity-fermion collisions. We have assumed that the states $\ket{\downarrow}_i$ effectively do not interact with the gas. This can be achieved, for example, by using a spin-dependent Feshbach resonance~\cite{chin_feshbach_2010} to tune the corresponding scattering length to a very large value. Note that in 1D the interaction strength is inversely proportional to the scattering length~\cite{olshanii_atomic_1998}.

We assume that the environment is initially in thermal equilibrium at temperature $T$, with $\hat{\rho}_T\propto e^{-\hat{H}_E/k_BT}$ the corresponding thermal state. The initial state of the system and environment is taken to be a tensor product
\begin{equation}
	\label{initial_product_state}
	\hat{\rho}(0) = \hat{\rho}_S(0) \otimes \hat{\rho}_T,
\end{equation}
which can be prepared since the $\ket{\downarrow}_i$ states do not perturb the gas. A specific experimental protocol to realise this is discussed in Sec.~\ref{subsec_ramsey}.

We want to infer the temperature of the gas by observing the dynamics of the probes. The latter are described by their reduced density matrix
\begin{equation}
	\label{rdm}
	\hat{\rho}_S(t) = \tr_E\left [e^{-i\hat{H}t/\hbar }\hat{\rho}(0)e^{i\hat{H}t/\hbar}\right ],
\end{equation}
obtained by tracing over the environment.  Since $[\hat{H}_I,\hat{H}_S] = 0$, the system Hamiltonian merely generates trivial phase factors that are irrelevant for the system-environment dynamics. From here on we remove these by working in a rotating frame via the transformation $\hat{\rho}_S(t) \to e^{i\hat{H}_S t/\hbar}\hat{\rho}_S(t) e^{-i\hat{H}_S t/\hbar}$, which is tantamount to setting $\varepsilon_i=0$. 

Let $\sigma$ label the different internal states of $S$, taking the values $\nolinebreak{\sigma \in \{\uparrow\uparrow, \uparrow\downarrow, \downarrow\uparrow, \downarrow\downarrow\}}$. It is straightforward to show that
\begin{equation}
	\label{rdm_elements}
 \mel{\sigma}{\hat{\rho}_S(t)}{\sigma'}  = \nu_{\sigma,\sigma'}(t) \mel{\sigma}{\hat{\rho}_S(0)}{\sigma'},
\end{equation}
where we have defined the complex functions
\begin{equation}
	\label{decoherence_functions}
	\nu_{\sigma,\sigma^\prime}(t) = \tr_E \left[e^{i\hat{H}_\sigma t/\hbar}e^{-i\hat{H}_{\sigma^\prime}t/\hbar}\hat{\rho}_T\right],
\end{equation}
with $\hat{H}_\sigma$ denoting the Hamiltonian of the environment conditioned on the internal state of the system:
\begin{equation}
    \hat{H}_\sigma = \mel{\sigma}{\hat{H}_E + \hat{H}_I}{\sigma}.
\end{equation}
The diagonal elements of $\hat{\rho}_S(t)$, with $\sigma=\sigma'$, are constant because the energy of the probes is conserved. The off-diagonal terms with $\sigma\neq \sigma'$ will evolve and generally decay in time according to $\nu_{\sigma,\sigma'}(t)$, which we refer to as the decoherence functions of our system. 

\subsection{Calculation of the decoherence functions}
\label{subsec_calcdec}

To calculate the decoherence functions in Eq.~\eqref{decoherence_functions}, we employ the Levitov formula or functional determinant approach (FDA)~\cite{abanin_fermi-edge_2005, dambrumenil_fermi_2005,schonhammer_full_2007}, which has been widely used to study nonequilibrium impurity systems~\cite{wang_functional_2022}. The FDA is a numerically exact method that maps a many-body expectation value into a determinant in single-particle space:
\begin{equation}
	\label{levitov}
	\nu_{\sigma,\sigma^\prime}(t) = \det[1- \hat{n} + \hat{n}e^{i\hat{h}_{\sigma^\prime}t/\hbar}e^{-i\hat{h}_{\sigma}t/\hbar}],
\end{equation}
This equation holds because Eq.~\eqref{decoherence_functions} involves only exponentials of quadratic fermionic operators. Here, $\hat{h}_\sigma$ is the single-particle equivalent of $\hat{H}_\sigma$, i.e. the Hamiltonian of a single particle in the gas conditioned on the impurities being in state $\sigma$. Meanwhile, $\hat{n}$ is an operator describing the initial Fermi-Dirac distribution of the gas atoms
\begin{equation}
	\hat{n} = \left[1+e^{(\hat{h}_{\downarrow\downarrow}-\mu)/k_BT}\right]^{-1}.
\end{equation}
We work in the grand canonical ensemble and use the chemical potential $\mu$ to fix the total number of atoms $N = \bar{n}L$, where $\bar{n}$ is the number density of the atoms. 

The box is modelled as an infinite square well with hard-wall boundary conditions at $x = \pm L/2$. The single-particle Hamiltonians $\hat{h}_\sigma$ are then differential operators of the form
\begin{equation}
	\label{h_sigma}
	\hat{h}_\sigma = -\frac{\hbar^2\nabla^2}{2m} + V_\sigma(x),
\end{equation}
defined on the interval $x\in [-L/2,L/2]$. Here, $V_\sigma(x)$ is the effective potential felt by the fermions when the impurities are in state $\sigma$. In the pseudopotential approximation, we have
\begin{align}
	\label{potentials}
	&  V_{\downarrow\downarrow}(x) = 0, \\
	&	V_{\uparrow\downarrow}(x) = -\frac{\hbar^2}{ma} \delta(x-x_1)\\
	&	V_{\downarrow\uparrow}(x) = -\frac{\hbar^2}{ma} \delta(x-x_2),\\
	& V_{\uparrow\uparrow}(x) = V_{\uparrow\downarrow}(x) + V_{\downarrow\uparrow}(x) .
\end{align}
From here on, we take the impurities to be placed symmetrically around the centre of the box, $x_1 = -x_0$ and $x_2 = +x_0$. This assumption can be made without loss of generality because we always take $L$ to be large enough to avoid boundary effects, so that only the impurity separation $\Delta x = x_2 - x_1 = 2x_0$ is relevant. Under this assumption, there are four independent decoherence functions that we need to fully describe the dynamics of the system. Further details on how we evaluate Eq.~\eqref{levitov} numerically can be found in Appendix~\ref{app:comp_details}.

\section{Dynamics of two impurities dephasing in a 1D Fermi gas}
\label{impurity_dynamics}

In this section we systematically investigate the dephasing dynamics of two impurities in a 1D Fermi gas. The physical scales of the gas are fully determined by the fermion number density, $\bar{n} = N/L$, and are given by the Fermi wavevector $k_F = \bar{n}\pi$,  the Fermi energy $E_F  = \hbar^2k_F^2/2m$,the Fermi time $\tau_F = \hbar/E_F$, the Fermi velocity $v_F = \hbar k_F/m$, and the Fermi temperature $T_F = E_F/k_B$. The interaction between the impurities and the gas is parameterized by the dimensionless parameter $k_F a$. We also define the interaction time between the impurities $\tau_i = \Delta x/v_F$. This can be understood as the time for excitations to travel from one impurity to the other, i.e.~the time after which bath-mediated interactions begin to play a role.

\begin{figure}
    \centering
    \includegraphics[width=0.49\textwidth]{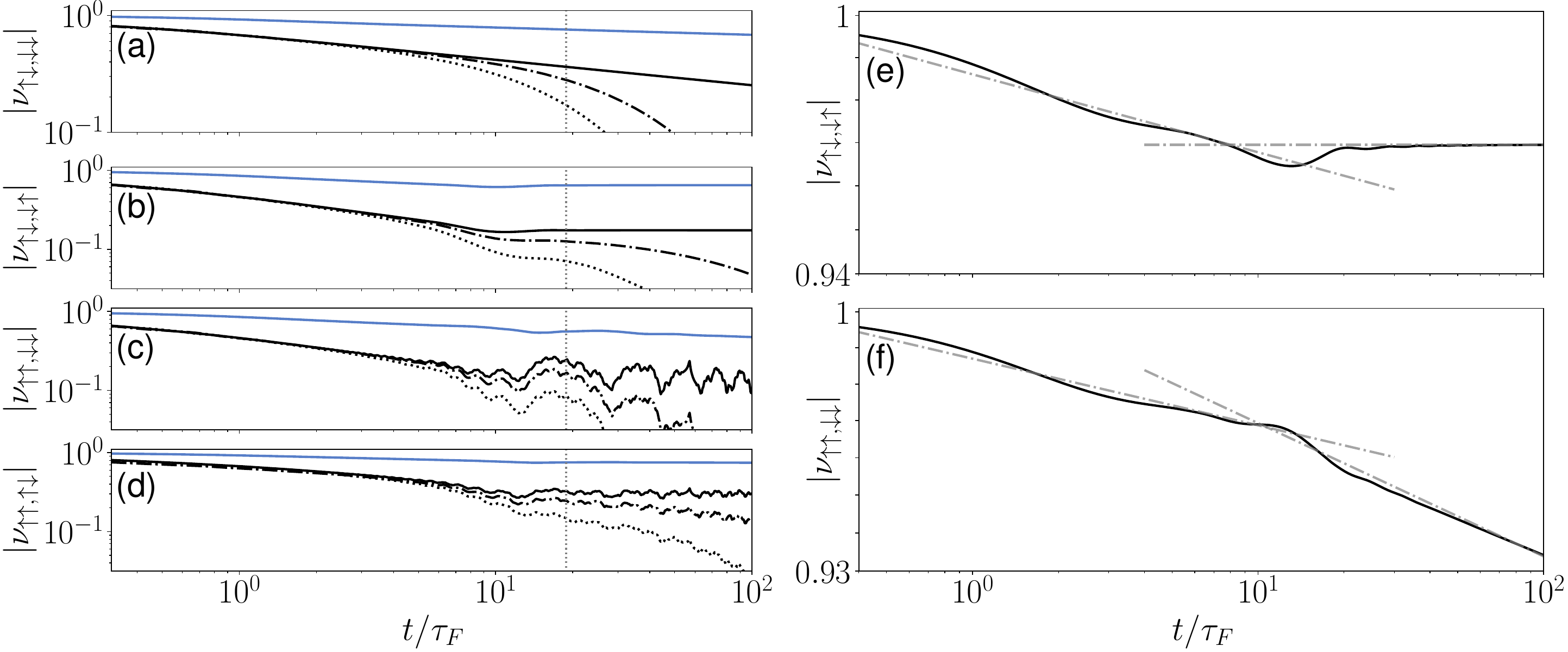}
    \caption{(a-d) The full dynamics of two impurity qubits with separation $k_F\Delta x/2\pi = 3$ coupled to a 1D fermionic bath at temperature $T = 0.0001T_F$ (solid), $T = 0.05T_F$ (dashdot) and $T = 0.1T_F$ (dotted) with coupling strength $k_Fa = -1$ (blue upper solid) and $k_Fa=-0.1$ (black lower). The Gray dotted line denotes the interaction time $\tau_i=\Delta x/v_F$. (e-f) The power law behaviour of $\nu_{\uparrow\uparrow,\downarrow\downarrow}$ and $\nu_{\uparrow\downarrow,\downarrow\uparrow}$ for $T = 0.0001T_F$ with an interaction strength of $k_Fa=-5$. The dotted lines are the power laws discussed in the main text valid for low temperature and weak coupling.}
    \label{dynamics_of_probes}
\end{figure}

In Fig.~\ref{dynamics_of_probes} we plot all four independent decoherence functions for both $k_F a = -0.1$ and $k_F a= -1$, which correspond to relatively strong and weak system-environment interaction strengths, respectively. For strong system-environment coupling we also plot the decoherence for temperatures $T/T_F = 0.1$, $0.05$ and $0.0001$. Here we focus exclusively on negative scattering lengths, in order to avoid the additional complication of the bound state that appears for $a>0$. While in general all four decoherence functions contribute to the dynamics, each one can be tied to a distinct initial state $\hat{\rho}_S(0)$, which is helpful to interpret their features.

We begin by examining the decoherence function $\nuuddd$ in Fig.~\ref{dynamics_of_probes}(a). This describes the situation where only a single impurity interacts with the gas. For example, given the initial condition $\hat{\rho}_S(0) = \ket{+}_1\!\bra{+}\otimes \ket{\downarrow}_2\!\bra{\downarrow}$ with
\begin{equation}
\label{plus_state}
\ket{+}_i = \frac{1}{\sqrt{2}}\left ( \ket{\uparrow}_i +\ket{\downarrow}_i \right ),
\end{equation}
the evolution of $\hat{\rho}_S(t)$ is determined purely by $\nuuddd(t)$ and all other matrix elements are constant or zero. The dynamics of a single heavy impurity qubit interacting with a Fermi gas has been extensively studied in the literature, e.g.~see~\cite{schmidt_universal_2018} for a comprehensive review of the three-dimensional case and the Supplemental Material of Ref.~\cite{mitchison_situ_2020} for analysis of the 1D case. 

Fig.~\ref{dynamics_of_probes}(a) reproduces the behaviour expected from these previous studies. For intermediate times between the Fermi time and the thermal timescale, $\tau_F\ll t\ll \hbar\beta$, the decoherence function has a universal power-law decay, with an exponent that depends on the interaction between the impurity and Fermi gas. For later times, thermal effects dominate and the decoherence function decays exponentially in time with a rate proportional to temperature. In order to elucidate this behaviour, in Appendix~\ref{app_cumulant} we derive an expression for the decoherence functions valid in the weak-coupling and low-temperature regime, using a cumulant expansion method adapted from Ref.~\cite{mitchison_situ_2020}. The result is that
\begin{equation}
	\label{weak_coupling_approx}
	|\nuuddd(t)|  \sim \begin{cases}
		t^{-\alpha} \quad &(\tau_F\ll t\ll \hbar\beta)\\ 
	e^{-\alpha t/\hbar \beta }  \quad & (\hbar\beta\ll t),
	\end{cases}
\end{equation}
where the exponents are determined by the dimensionless coupling strength
\begin{equation}
	\alpha = \left(\pi k_F a\right)^{-2}.
\end{equation}

The decoherence functions $\nuuddu$ and $\nuuudd$ are shown in Fig. \ref{dynamics_of_probes}(b) and (c). These are the relevant decoherence functions for initial Bell state preparations $\hat{\rho}_S(0) = \dyad{\Psi^{\pm}}$ and $\hat{\rho}_S(0) = \dyad{\Phi^{\pm}}$, respectively, where
\begin{equation}
	\begin{split}
		|\Psi^\pm\rangle &= \frac{1}{\sqrt{2}}\left( \ket{\uparrow_1\downarrow_2} \pm \ket{\downarrow_1 \uparrow_2}\right)\\
		|\Phi^\pm\rangle &= \frac{1}{\sqrt{2}}\left(\ket{\uparrow_1\uparrow_2} \pm \ket{\downarrow_1\downarrow_2} \right).
	\end{split}
\end{equation}
 For short times, $t\ll \tau_i$, both decoherence functions have the same power-law behavior as two impurities in independent baths, i.e.~ $\nu_{\sigma,\sigma'}\sim t^{-2\alpha}$. This reflects the fact that, for times much less than the bath-mediated interaction time, the two impurities should evolve independently. This implies that, for $t<\tau_i$, we have
\begin{equation}
	\label{product_channel}
    \hat{\rho}_S(t) = \mathcal{E} \hat{\rho}_S(0) \approx \left (\mathcal{E}_1 \otimes \mathcal{E}_1\right )\hat{\rho}_S(0),
\end{equation}
where $\mathcal{E}$ is the quantum channel describing the exact two-impurity evolution and $\mathcal{E}_1$ is the channel describing a single impurity immersed in a Fermi gas. Eq.~\eqref{product_channel} immediately implies that  $\nuuddu\approx |\nuuddd|^2$ and $\nuuudd\approx \nuuddd^2$. Therefore, the phase of $\nuuddu$ vanishes while the phase of $\nuuudd$ is twice that of a single impurity, and their magnitudes are identical. However, around the interaction time $\tau_i$ these two decoherence functions begin to show drastically different behaviour: the decay of $\nuuddu$ slows down completely, while $\nuuudd$ begins to show marked oscillations with approximate period $\tau_i$. These oscillations represent a non-Markovian effect, where excitations from one impurity travel through the gas and hit the other one, representing backflow of information from the environment into the system~\cite{rivas_quantum_2014, breuer_colloquium_2016}. 

These dynamical features for $t\lesssim\tau_i$ are qualitatively captured by the cumulant expansion derived in Appendix~\ref{app_cumulant}. For short times $t\ll \tau_i$, our analytical theory predicts that both $\nuuddu$ and $\nuuudd$ behave as $\nu_{\sigma,\sigma'}(t) \sim t^{-2\alpha}$, as expected for independent impurities. After the impurity interaction time $t = \tau_i$, we find that the algebraic decay changes exponent as
\begin{equation}
	\label{new_exponent}
	\alpha \to \alpha\left(1\pm \cos^2(2k_F\Delta x)\right),
\end{equation}
where the plus and minus signs correspond to $\nuuudd$ and $\nuuddu$, respectively. A comparison between this analytical prediction and the exact numerics can be seen in Fig. \ref{dynamics_of_probes} (e) and (f), demonstrating excellent agreement within the weak-coupling and low-temperature regime where the cumulant expansion is valid. 

  The change of exponent at $t=\tau_i$ manifests the well-known phenomenon of super- and sub-decoherence~ \cite{palma_quantum_1996,reina_decoherence_2002, klein_dynamics_2007, kattemolle_conditions_2020}. In Appendix~\ref{app_cumulant} we provide an explanation for this effect: the Bell states $\ket{\Phi^{\pm}}$ and $\ket{\Psi^{\pm}}$ couple respectively to density fluctuations at the impurity positions that are in phase or anti-phase, respectively, which have very different low-frequency properties. Moreover, in Eq.~\eqref{new_exponent} we recognise the sinusoidal spatial dependence that typically arises in perturbed Fermi gases, with the same period of $\pi/k_F$ that characterises both Friedel oscillations and the RKKY interaction. Therefore, the transition from normal to sub- or super-decoherent dynamics manifests the time-retarded effect of RKKY-like correlations developing in real-time. Note that the sub- and super-decoherent behaviour persists only up to the thermal time, $t\simeq \hbar\beta$, after which both decoherence functions decay exponentially as $\sim e^{-2\alpha t/\hbar\beta}$. This is the decay expected for two independent impurities, meaning that thermal noise washes out the dynamical effect of RKKY-like interactions on the decoherence signal at long times. We note that RKKY-like interactions have been observed in ultracold atomic gases recently~\cite{edri_observation_2020}.

Finally, Fig. \ref{dynamics_of_probes} d shows the decoherence function $|\nuuuud|$. This is the relevant decoherence function for the initial state $\hat{\rho}_S(0) = \ket{\uparrow}_1\!\bra{\uparrow}\otimes \ket{+}_2\!\bra{+}$, which describes a situation where the pseudospin of the first impurity is flipped at the same time that the second impurity is prepared in a superposition. Thus, for short times, $\nuuuud$ is indistinguishable from $\nuuddd$. For times $t\gtrsim \tau_i$ the perturbations from the first impurity hit the second one, causing the decoherence to slow down. The cumulant expansion at second order fails to capture this behaviour, indicating that this is a non-perturbative effect.

Aside from the temperature dependence of the long-time exponential decay, the interaction effects discussed above depend strongly on both temperature and coupling strength. In particular, non-Markovian effects due to exchange of excitations are most prominent at low temperatures, and disappear as the temperature increases. Observing these features thus yields useful information on temperature, as we show in the next section. 

\section{Thermometry with correlated dephasing probes}
\label{thermometry}

\subsection{Local temperature estimation theory}
\label{quant_est}

Since the impurity dynamics depends on the temperature of the gas, we can infer the temperature from measurements of many identical preparations of the probes. We first consider a general ``prepare and measure'' scenario: a given probe state $\hat{\rho}_S(0)$ is prepared as in Eq.~\eqref{initial_product_state}, the system decoheres for a time $t$, and then a measurement in a given basis is made on the resulting state $\hat{\rho}_S(t)$. Repeating this procedure for $M$ identical preparations yields a sequence of measurement outcomes $\boldsymbol x = \{x_1, x_2, \dots, x_M\}$. In the most general case, the measurement may be a positive operator-valued measure (POVM): a collection of positive operators $\hat{\Pi}(x)>0$ that are normalised as $\sum_x \hat{\Pi}(x) = 1$.

Using our knowledge of the temperature-dependence of $\hat{\rho}_S(t)$, a temperature estimate $\check{T}(\textbf{x})$ can be constructed from the measurement data $\textbf{x}$, e.g.~using maximum-likelihood estimation. Any such estimate will carry uncertainty due to the randomness inherent to quantum measurements and the finite number of samples $M$. The achievable precision depends not only on the probe state $\hat{\rho}_S(T)$ but also the specific choice of POVM $\{\hat{\Pi}(x)\}$. 

To quantify the precision attainable in our setup, we use the theory of local quantum parameter estimation~\cite{paris_quantum_2009}. Here, the central quantity is the quantum Fisher information (QFI), $\mathcal{F}_T$, which provides a lower bound on the variance of any unbiased estimator. In particular, an unbiased estimator obeys $\mathbb{E}[\check{T}(\textbf{x})] = T$, while its variance obeys the (quantum) Cram\'{e}r-Rao bound~\cite{braunstein_statistical_1994}
\begin{equation}
	\label{QCRB}
 \mathbb{E} \left [\left (\check{T}(\textbf{x}) - T\right )^2\right ] \geq \frac{1}{MF_T}\geq \frac{1}{M\mathcal{F}_T}.
\end{equation}
Here, $F_T$ is the Fisher information for the measurement $\{\hat{\Pi}(x)\}$,
\begin{equation}
	F_T\left (\hat{\rho}_S(t),\{\hat{\Pi}(x)\}\right ) =\sum_x p(x) \left(\pdv{\ln{p(x)}}{T}\right)^2,
\end{equation}
where $p(x) = \tr\left [ \hat{\rho}_S(t) \hat{\Pi}(x)\right ]$ is the probability of obtaining outcome $x$. The QFI is defined as maximum of the Fisher information over all POVMs~\cite{braunstein_statistical_1994},
\begin{equation}
	\label{Fisher_maximum}
	\mathcal{F}_T\left (\hat{\rho}_S(t)\right ) = \max_{\{\hat{{\Pi}(x)}\}} 	F_T\left (\hat{\rho}_S(t),\{\hat{\Pi}(x)\}\right ).
\end{equation}
Therefore, the QFI quantifies the maximum information on temperature that can be obtained from repeated measurements on the state $\hat{\rho}_S(t)$. 

The maximum in Eq.~\eqref{Fisher_maximum} is achieved by projective measurements of the symmetric logarithmic derivative (SLD), $\hat{\Lambda}_T$. Writing the probe state in its eigenbasis as $\hat{\rho}_S(t) = \sum_n r_n \ket{r_n}\bra{r_n}$, the QFI and SLD are given explicitly by~\cite{paris_quantum_2009}
\begin{align}
	\label{QFI}
&	\mathcal{F}_T = 2\sum_{m,n} \frac{\lvert\bra{r_m} \partial_T\hat{\rho}_S \ket{r_n}\rvert^2}{r_m+r_n},\\
	\label{SLD}
	&\hat{\Lambda}_T = 2\sum_{m,n} \frac{\bra{r_m} \partial_T\hat{\rho}_S \ket{r_n}}{r_m+r_n} \ket{r_m}\bra{r_n}.
\end{align}
Note that the SLD is explicitly dependent on temperature, and the quantum Cram\'er-Rao bound~\eqref{QCRB} can only be saturated in the asymptotic limit of $M\gg 1$. Therefore, local estimation theory is most relevant when some coarse information on temperature is already known and a large quantity of measurement data is available.  Alternative approaches to temperature estimation based on Bayesian statistics~\cite{rubio_global_2021,mok_optimal_2021,boeyens_uninformed_2021,jorgensen_bayesian_2022,mehboudi_fundamental_2022,alves_bayesian_2022} and single-shot information theory~\cite{meyer_quantum_2023} have recently been developed, allowing thermometry even with few measurements and no prior information. Nevertheless, the QFI is very useful as a quantitative benchmark for our impurity thermometer in different parameter regimes. 
%In the following, we focus mainly on the quantum signal-to-noise-ratio (QSNR), defined by
%\begin{equation}
%    \mathcal{Q}^2 = T^2\mathcal{F}_T,
%\end{equation}
%which bounds the fractional sensitivity as
%\begin{equation}
%    \frac{T}{\Delta T}\leq \sqrt{M}\mathcal{Q}. 
%\end{equation}

\subsection{Ramsey interferometry protocol}
\label{subsec_ramsey}

Since the impurities become correlated through their interaction with the gas, the SLD is generally a non-local observable that may be difficult to measure. We therefore consider the following thermometry protocol based on Ramsey interferometry, which is feasible in cold-atom systems~\cite{cetina_ultrafast_2016,adam_coherent_2022}. To generate the initial product state~\eqref{initial_product_state}, both of the impurities start in the non-interacting state $\ket{\downarrow}_i$ while the gas is prepared in a thermal state. The impurities may either be left \textit{in situ} during this preparation phase~\cite{cetina_ultrafast_2016}, or transported into the Fermi gas by a moving trap potential~\cite{schmidt_quantum_2018,adam_coherent_2022}. At $t=0$, a  $\pi/2$ pulse is applied to the impurities, generating the initial state
\begin{equation}
	\label{initiial_plus_state}
	\hat{\rho}_S(0) = \ket{+}_1\!\bra{+} \otimes \ket{+}_2\!\bra{+}.
\end{equation}
The system is then left to decohere in contact with the gas for a time $t$, giving rise to a state $\hat{\rho}_S(t)$ which is a function of all four decoherence functions discussed in Sec.~\ref{impurity_dynamics}. Finally, another $\pi/2$ pulse is applied with a phase $\phi$ relative to the first pulse, and then the population of the qubit energy eigenstates is immediately measured projectively, e.g.~by using laser pumping to induce state-dependent flourescence. 

In experiments with multiple impurities embedded in ultracold atomic gases, it is typically not possible to address the individual impurities during the measurement. Instead, only the total measurement signal is accessible, e.g.~in Ref.~\cite{adam_coherent_2022} the total number of impurities left in the ground state after the final $\pi/2$-pulse is counted. This is equivalent to measuring the expectation value of the observable
\begin{equation}
\label{local_measurement}
    \hat{O}(\phi) = \sum_{i=1}^2\left[\cos(\phi)\hat{\sigma}_{i}^{x} + \sin(\phi)\hat{\sigma}_{i}^{y}\right].
\end{equation}
The temperature uncertainty $\Delta T$ expected for $M\gg 1$ measurements of this observable can be quantified by the error propagation formula~\cite{toth_quantum_2014}
\begin{equation}
	\label{precision_non_optimal}
\Delta T = \frac{\Delta \hat{O}}{ \sqrt{M} |\partial_T \langle\hat{O}\rangle| },
\end{equation}
where $\Delta \hat{O}^2 = \langle \hat{O}^2\rangle - \langle \hat{O}\rangle^2$ is the operator variance and $\langle \hat{O}\rangle$ is the expectation value in the state $\hat{\rho}_S(t)$. The signal-to-noise ratio is then
\begin{equation}
	\label{SNR}
 	\frac{T}{\Delta T}  =  \sqrt{M}  \frac{ T |\partial_T \langle\hat{O}\rangle|}{ \Delta \hat{O}} \equiv \sqrt{M} \mathcal{S}_T({\hat{O}})
\end{equation}
which defines the effective temperature sensitivity $\mathcal{S}_T({\hat{O}})$ for measurements of the observable $\hat{O}$. 

\begin{figure*}
	\centering
	\includegraphics[width=0.99\textwidth]{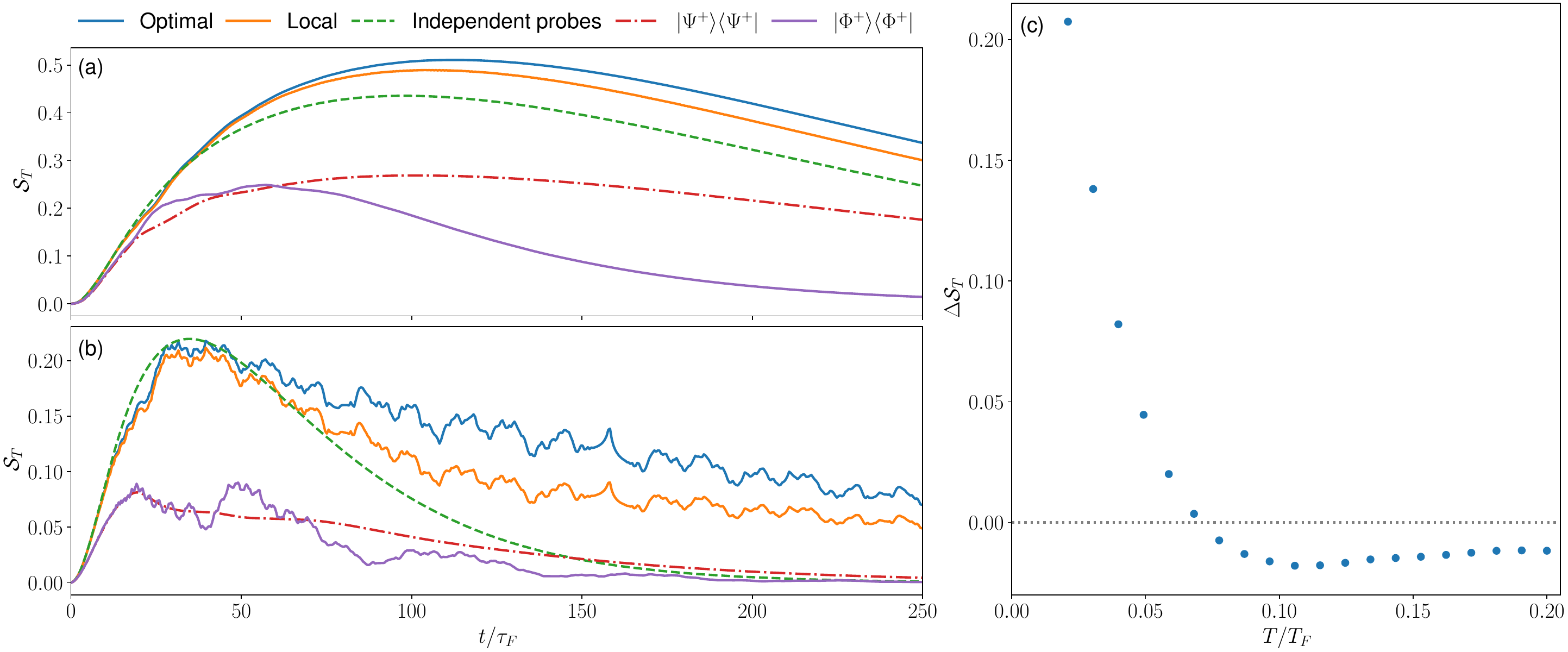}
	\caption{(a) and (b) The signal-to-noise ratio of different measurements for the two impurity qubits separated by $k_F\Delta x/2\pi = 5$, temperature $T=0.05T_F$, and interaction strength $k_F a = -1$ (a) and $k_F a = -0.1$ (b). Measurement of the SLD including bath-induced interactions(upper blue solid lines), the SLD of two independent impurities (green dashed lines), the operator of Eq. (\ref{precision_non_optimal}) with angle $\phi$ that gives the maximum information (middle orange solid line). We also investigated the effect of starting in one of the Bell states $\dyad*{\Psi^+}$ or $\dyad*{\Phi^+}$ and doing measurement on their respective SLD (lower red dash-dotted and purple solid lines). (c) The difference between maximal sensitivity with and without bath-induced interactions (see Eq.~(\ref{delta_S})) for different temperatures.}
	\label{qsnr_results}
\end{figure*}

The sensitivity $\mathcal{S}_T$ will be our metric of performance in the following. It can be shown that the sensitivity is maximised by measurements of the SLD because~\cite{mehboudi_using_2019}
\begin{equation}
	\label{QSNR}
	\mathcal{S}_T(\hat{\Lambda}_T) = T \sqrt{\mathcal{F}_T},
\end{equation}
in which case, according to Eq.~\eqref{SNR}, the temperature error $\Delta T$ asymptotically saturates the quantum Cram\'er-Rao bound~\eqref{QCRB}. For local observables of the form in Eq.~\eqref{local_measurement}, the sensitivity is generally smaller; however, for any state $\hat{\rho}_S(t)$ we can find the optimal operator $\hat{O}(\phi)$ to measure by maximising $\mathcal{S}_T$ over $\phi$. Like the SLD, the optimal $\phi$ depends on $T$ and thus some prior knowledge about the temperature is needed for this to work in practice.

\subsection{Thermometric performance and collective advantages}
\label{thermometric_performance}

We now discuss how the dynamical features explored in Sec.~\ref{impurity_dynamics} affect the performance of our two-qubit thermometer. The blue curves in Fig. \ref{qsnr_results}  (a) and (b) show the optimal sensitivity, $\mathcal{S}_T(\hat{\Lambda}_T) = T\sqrt{\mathcal{F}_T}$, as a function of time in various different scenarios for temperature $T = 0.05T_F$ and two interaction strengths: $k_F a = -1$ [Fig.~\ref{qsnr_results}(a)] and $k_F a = -0.1$ [Fig.~\ref{qsnr_results}(b)]. The sensitivity peaks at a particular moment in time, which defines the optimal time to perform the measurement. We see that the peak sensitivity is generally larger at weak system-environment coupling but also occurs at a later time. A similar tradeoff between sensitivity and measurement time was found in previous work on thermometry with single impurities~\cite{mitchison_situ_2020}. At stronger coupling, the optimal sensitivity exhibits complicated oscillations. These are a consequence of non-Markovian effects induced by the impurities exchanging excitations via the gas, as discussed in Sec.~\ref{impurity_dynamics}.

An interesting question is whether bath-induced interactions improve the precision of our two-impurity thermometer. In order to assess this, we compare to the case of two independent impurities with $\hat{\rho}_S(t) = [\hat{\rho}_1(t)]^{\otimes 2}$, where $\hat{\rho}_1(t)$ is the state of a single impurity dephasing in a 1D Fermi gas. That is, the diagonal elements of $\hat{\rho}_1(t)$ are constant while the off-diagonal elements are proportional to  $\nuuddd(t)$. This is equivalent to the situation where the impurity separation is very large, so that $\tau_i \gg \hbar\beta$ and thermal decoherence kick in before interactions can play any role. The corresponding QFI is additive, $\mathcal{F}_T([\hat{\rho}_1(t)]^{\otimes 2}) = 2\mathcal{F}_T(\hat{\rho}_1)$. Fig.~\ref{qsnr_results}(a) shows that the optimal sensitivity of two impurities in a common bath can be markedly larger than the corresponding value for two independent impurities (green dashed curves in Fig.~\ref{qsnr_results} (a) and (b)). This effect, which only arises for sufficiently weak coupling and low temperature, indicates that bath-induced interactions yield an advantage for thermometry. 

Remarkably, this advantage persists even when restricted to more realistic local observables as in Eq.~\eqref{local_measurement}. To show this, we numerically optimize the angle $\phi$ to find the observable that gives the largest sensitivity $\mathcal{S}_T$ at each point in time. The results, shown by the orange curve in Fig.~\ref{qsnr_results} (a) and (b), demonstrate that optimal local measurements may yield almost as much temperature information as measurements of the SLD. This is true for a wide range of parameters. Similar effects have been observed previously in the context of equilibrium thermometry \cite{campbell_global_2017}.

 To quantify how the collective thermometry advantage depends on temperature, we compute the difference in peak thermometric performance between the independent impurities and impurities with bath-induced interactions
\begin{equation}
\label{delta_S}
    \frac{\Delta \mathcal{S}_T}{T} = \max_t \sqrt{\mathcal{F}_T[\hat{\rho}_S(t)]} - \max_t \sqrt{2\mathcal{F}_T[\hat{\rho}_1(t)]}.
\end{equation}
Here, as above, $\hat{\rho}_S(t)$ is the state of two impurities immersed in a common environment, whereas $\hat{\rho}_1(t)$ is the state of a single impurity dephasing in a 1D Fermi gas. As shown in Fig.~\ref{qsnr_results} (c), we observe a large and positive $\Delta \mathcal{S}_T$ at low temperatures, signifying a significant collective advantage. However, at higher temperatures this trend reverses and we find that independent impurities yield higher thermometric precision. At even higher temperatures, the difference goes to zero because thermal fluctuations wipe out any bath-induced interactions. Similarly, we have observed the collective advantage to be most prominent at weak coupling. In all the cases we have investigated, sufficiently strong coupling led to independent impurity probes outperforming the correlated probes. An example of this can be seen in Fig.~\ref{qsnr_results} (b). 

\begin{figure}[b]
	\centering
	\includegraphics[width=0.49\textwidth]{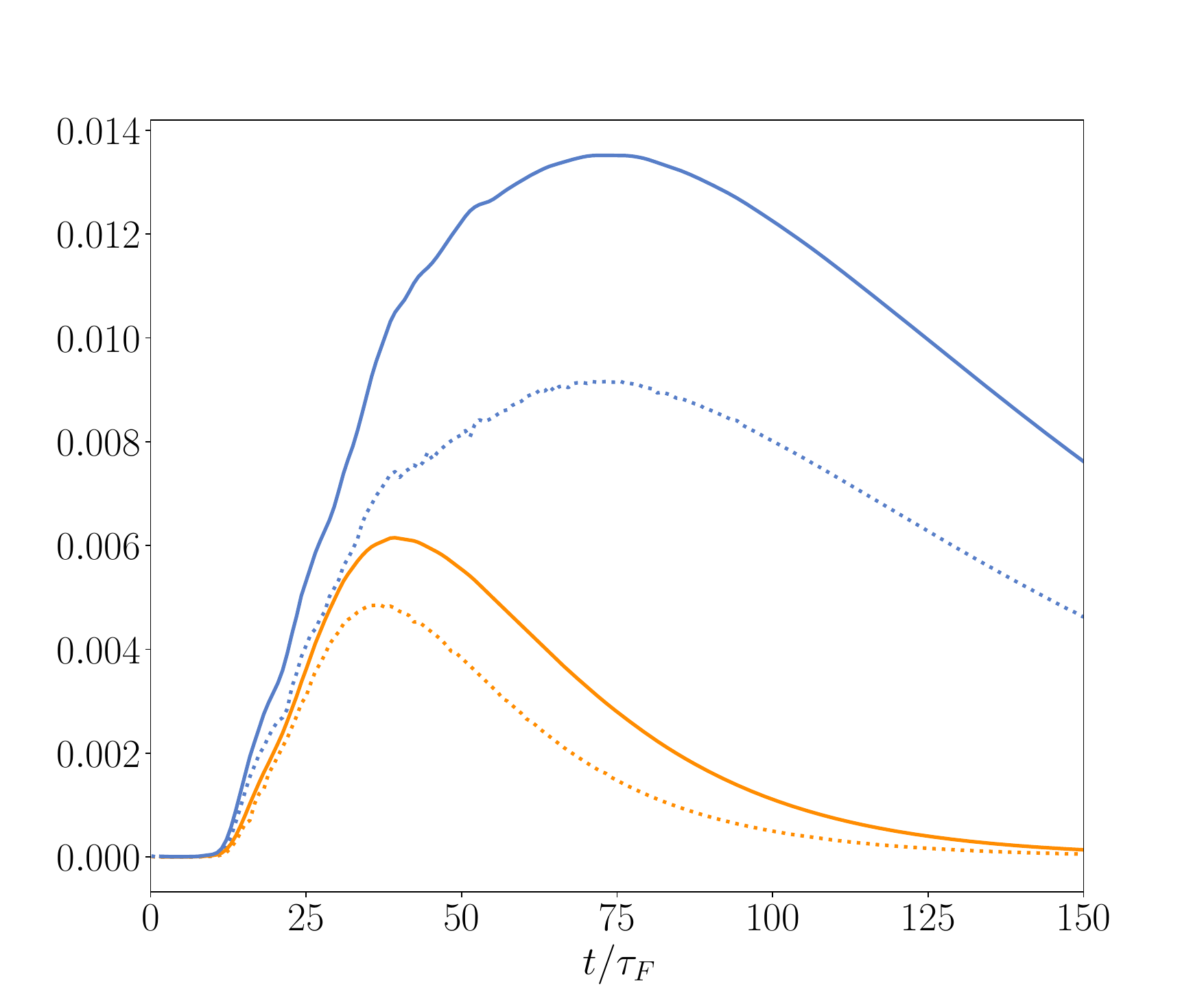}
	\caption{The mutual information (solid lines) and quantum discord (dotted lines) between the two impurity probes as a function of time, with separation $k_F\Delta x/2\pi = 5$, interaction strength $k_Fa = -1$ and temperatures $T=0.05T_F$ (upper blue lines) and $T=0.125T_F$ (lower orange lines).}
	\label{fig:mutual_info}
\end{figure}

To further elucidate the role of bath-induced interactions, we quantify the correlations that develop between the two impurity probes using the quantum mutual information
\begin{equation}
	\label{mutual_info}
	I(1:2) = S(\hat{\rho}_1) + S(\hat{\rho}_2) - S(\hat{\rho}_S),
\end{equation}
where $\hat{\rho}_i$ is the reduced density matrix for impurity $i$, e.g.~$\hat{\rho}_1 = \tr_2 (\hat{\rho}_S)$, and $S(\hat{\rho})$ is the von Neumann entropy
\begin{equation}
	S(\hat{\rho}) = -\tr( \hat{\rho}\ln \hat{\rho}).
\end{equation}
 We also investigate the amount of correlation that is purely quantum in nature by means of the quantum discord~\cite{ollivier2001quantum, Henderson_2001}. The quantum discord the difference between the mutual information and the purely classical correlations:
\begin{equation}
    \mathcal{D}(1:2) = I(1:2) - \mathcal{C}(1:2).
\end{equation}
Further details on the quantum discord, and the precise definition of the classical correlations $\mathcal{C}(1:2)$, are given in Appendix \ref{app:quant_corr} (and see Ref.~\cite{Adesso_2016} for a comprehensive review). Fig.~\ref{fig:mutual_info} shows the mutual information and quantum discord of the two probes as a function of time. We see that correlations, both quantum and classical, are strongest in the low-temperature regime. We also find that the correlations between the impurities are much weaker in the regime of strong impurity-gas coupling (not shown in figure). Conversely, for high temperature or strong coupling, the correlations quickly vanish. 

However, correlations alone are not sufficient to obtain a precision advantage, as the following example demonstrates. Consider preparing the impurities in the maximally entangled Bell states $\hat{\rho}_S(0) = \dyad{\Phi^+}$ or $\hat{\rho}_S(0) = \dyad{\Psi^+}$. Entangled initial states are known to give a precision boost in many metrological settings, e.g. for phase estimation in cold atom systems~\cite{meyer_experimental_2001,schleier-smith_states_2010}. For our system, however, this is not the case, as shown by the purple and red curves in Fig.~\ref{qsnr_results}. For the $\ket{\Psi^+}$ state, this can be understood because the relevant decoherence function is almost purely real, while much of the temperature information at weak coupling is known to be contained in bath-induced phase shifts~\cite{mitchison_situ_2020}. For the initial $|\Phi^+\rangle$ state, phase information is amplified but the super-decoherence effect discussed in Sec.~\ref{impurity_dynamics} causes the matrix elements of $\hat{\rho}_S$ to decay too quickly to take advantage of this effect. 

In summary, correlations between the impurities are generated by bath-mediated interactions, and these correlations can increase the temperature sensitivity. This effect is strongest at weak coupling and low temperature and, most interestingly, can be exploited using only product-state preparations and local measurements, e.g.~via Ramsey interferometry. However, since these bath-mediated interactions vary rapidly with the distance between the impurities, the impurity's positions must be controlled with a precision comparable to the Fermi wavelength [c.f. Eq.~\eqref{new_exponent}] to take advantage of these collective effects. Moreover, finding the optimal measurement and constructing the estimator requires solving the dynamical problem for two impurities, which is significantly more complicated than the single-impurity case. In the following section, therefore, we consider whether one may ignore bath-induced correlations and still obtain a reasonable temperature estimate. 

\subsection{When can we assume uncorrelated probes?}
\label{subsec_indep_probes}

In principle, our functional determinant approach can be scaled to describe the dynamics of an arbitrary number of probes, $M$. In practice, however, this becomes impractical due to the large number of different decoherence functions that must be evaluated, as well as numerical instabilities and convergence issues that must be carefully addressed (see Appendix~\ref{app:comp_details}). This motivates us to ask: under what conditions the bath-induced correlations can be ignored without sacrificing the quality of the temperature estimate? This would enable one to treat the impurities as independent, so that only the dynamics of a single impurity need be solved, which is far simpler. 

To investigate the error arising in the temperature estimate by assuming the two impurity probes are uncorrelated, we adopt the following procedure:
\begin{enumerate}
  \item Numerically find the operator $\hat{O}_{\rm max}(\phi)$ that gives the most information on temperature as well as the optimal time $t_{\rm max}$ to perform the measurement, assuming the impurities are evolving in independent Fermi gases of temperature $T$. This yields the optimum measurement that an experimenter who is unaware of bath-induced interactions would perform. 
  \item Evaluate the reduced density matrix for two impurities in a shared bath of temperature $T$ and evaluate the expectation value $\langle \hat{O}_{\rm max}\rangle = \tr \left (\hat{\rho}_S(t_{\rm max})\hat{O}_{\rm max}\right )$. This yields the expected result that the unaware experimentalist would measure. 
  \item Find $T_{\rm est}$, the temperature that the unaware experimenter would infer from the result $\langle \hat{O}_{\rm max}\rangle$ (assuming $M\to \infty$ measurements), using the model of independent impurities. 
  \item Compare $T_{\rm est}$ with $T$ to infer the error that would be incurred.
\end{enumerate}

The error in assuming uncorrelated probes depends crucially on the dynamics of the system at the measurement time $t_{\rm max}$. If $t_{\rm max} \ll \tau_i$, the impurities remain uncorrelated and $T_{\rm est} = T$ to an excellent approximation. Therefore, we instead focus on impurities with a separation such that the interaction time is less than or similar to the $t_{max}$, where the effect of bath-induced interactions on the temperature estimate is most significant. 

\begin{figure}
	\centering \includegraphics[width=0.49\textwidth]{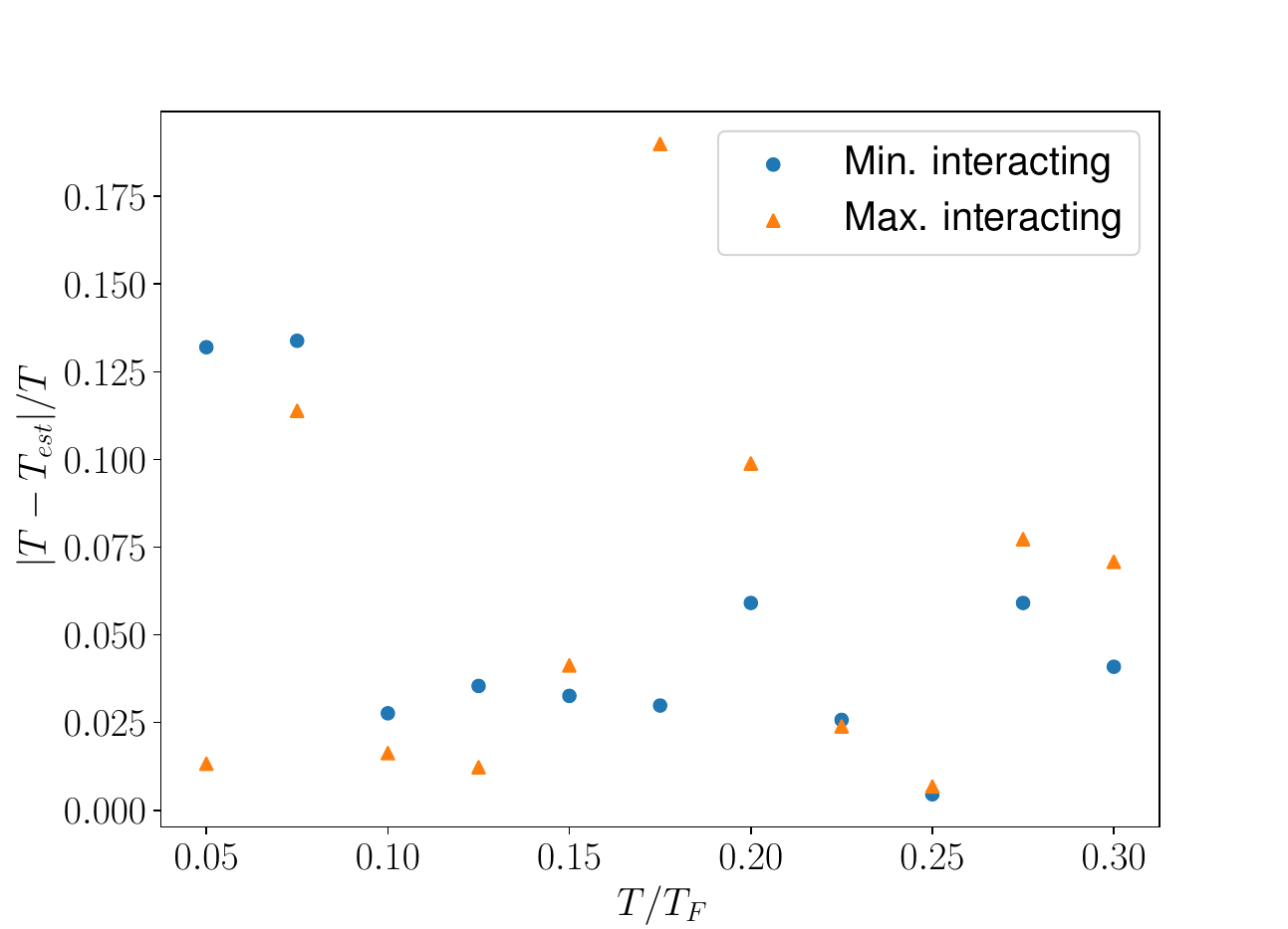}
	\caption{The relative error when assuming uncorrelated impurity probes as a function of temperature for coupling strength $k_F a = -1$. The blue and orange points are for separations such that $k_F\Delta x = n\pi$ and $k_F\Delta x = n\pi/2$, respectively. }
	\label{how_wrong}
\end{figure}

Fig.~\ref{how_wrong} shows the error in assuming uncorrelated probes for different temperatures for two separations of the impurities. From Section \ref{impurity_dynamics}, we know the effect of the bath-induced impurity interaction is quantified by $\cos^2(2k_F\Delta x)$. We consider the two extremal cases $2k_F\Delta x = n\pi$ and $2k_F\Delta x = n\pi/2$ ($n\in \mathbb{Z}$), where this interaction is maximal or minimal, respectively. We find that in both cases, and for most temperatures considered, approximating the impurities as independent yields a relative error $|T-T_{\rm est}|/T \lesssim 10\%$.  However, the error depends in a complicated way on the temperature, because this determines the optimal measurement time $t_{\rm max}$. For example, at temperature $T = 0.175T_F$ the error for the maximally interacting probes is large because $t_{\rm max}$ is close to an integer multiple of $\tau_i$.

These results suggest that, at least for two fixed impurities, achieving the best precision requires an estimator that explicitly accounts for bath-induced correlations. However, one may hope that, when averaging over the signal from many impurities at widely varying positions, the effect of correlations averages out. This presumably depends on the impurities' spatial distribution, but we leave a careful analysis of this problem for $M>2$ to future work.

\section{Discussion \& conclusions}
\label{sec_discussion}

In this work, we have proposed and analysed a thermometry protocol based on the dynamics of two qubit impurities dephasing in a 1D gas of ultracold fermions. We have solved the quantum evolution of the two impurities exactly, including the effect of bath-induced interactions. We have also gained valuable physical intuition into the observed behaviour at different timescales by means of a perturbative cumulant expansion. We have found that certain impurity decoherence functions manifest a retarded RKKY-like interaction, giving rise to sub- and super-decoherent behaviour depending on the initial state, which persists for intermediate times $\tau_i<t<\hbar \beta$ and is eventually washed away by thermal fluctuations. 

To understand how the bath-mediated interactions affect the achievable precision, we have compared the thermal sensitivity of our two-qubit thermometer to a pair of impurities interacting with independent environments. We have found that, at low temperatures and weak coupling, bath-induced correlations between the impurities can enhance precision. These results reinforce other recent work showing that bath-mediated interactions can be helpful in the context of low-temperature thermometry~\cite{gebbia_two-qubit_2020, planella_bath-induced_2022,brenes_multi-spin_2023}. This conclusion is by no means obvious, since correlations between the impurities could also arise from redundant encoding of temperature information, thus reducing the signal-to-noise ratio relative to truly independent measurements~\cite{hangleiter_nondestructive_2015,mitchison_probing_2016}. 

In order to exploit these correlations to the full, one would need to measure the non-local SLD observable. However, we have shown that a simple Ramsey protocol can approach the optimal precision, without the need to individually address the impurities or perform entangling operations. Moreover, we have quantified the systematic error incurred by neglecting correlations altogether. Our results show that this error depends strongly on the impurity separation, but remains on the order of a few percent for most system parameters at not-too-low temperatures. This suggests that it may be an acceptable approximation to neglect bath-induced correlations in most situations, potentially simplifying future thermometry experiments with many impurities embedded in a single copy of the gas. 

We have found that bath-mediated correlations can yield a thermometric advantage at weak coupling and low temperature when both classical and quantum correlations are allowed to build up between the impurities. These correlations are temperature dependent and thus contain more information on temperature than would be available for independent impurities, thus allowing for increased precision. However, our analysis also shows that correlations can be detrimental to thermometric precision. Indeed, the loss of precision in the presence of correlations has also been shown for classical processes \cite{Radaelli_2023}. The question of when correlations are helpful or harmful to precision would be an interesting starting point for further studies. 

It is possible to generalize the techniques presented in this paper to investigate other systems, for example higher-dimensional systems. It has recently been suggested that the precision one is able to obtain depends crucially on the spatial dimensionality of the system~\cite{khan_subnanokelvin_2022} and that thermometry in higher dimensions could be more effective. However, we also expect bath-induced interactions to be less prominent in higher dimensions because the excitations from the impurity-bath interactions will spread out more. It would be interesting to investigate how the dynamics, as well as the thermometric performance, would change for several impurities decohering in a higher-dimensional Fermi gas. 

In our current work we have only considered a non-interacting environment. An interesting question is then whether bath-mediated correlations will still be useful in the presence of interactions between the bath atoms.  If the environment is two- or three-dimensional, we expect Fermi liquid theory to apply (for repulsive interactions at low temperature). The environment will then behave as a collection of non-interacting fermions and our discussion above on higher-dimensional systems should still hold. If the interaction is attractive, we expect the environment to behave as a BCS Fermi superfluid, and it has been shown that the FDA can be straightforwardly generalized to this case \cite{wang_exact_2022,wang_heavy_2022}. In the case of an environment of interacting fermions in 1D, we expect the environment to behave as a Luttinger liquid at low temperatures, with low-energy excitations that are bosonic. Several works have investigated the use of impurity dynamics for thermometry of one-dimensional  bosonic superfluids~\cite{khan_subnanokelvin_2022, mehboudi_using_2019,yuan_quantum_2023}, and it has been shown that bath-mediated interactions can enhance thermometric precision~\cite{planella_bath-induced_2022}. 

Our approach could also be adapted to explore thermometry in the presence of different confining potentials \cite{knap_time-dependent_2012} or charged impurities~\cite{oghittu_quantum-limited_2022}. Other properties of these systems beyond temperature, such as transport coefficients~\cite{Saha2021}, could also be extracted using correlated impurities. Moreover, it has been suggested that using multi-dimensional spectroscopy instead of the Ramsey protocol considered in this text could yield extra information \cite{wang_two-dimensional_2022,wang_multidimensional_2023}, and the procedure presented here could be generalised to such scenarios as well.

Finally, it would be interesting to see how thermometric performance scales with the number of impurities $M$, especially for $M \gg 1$. A closely related question is how disturbing such a temperature measurement would be in terms of heat absorbed by the environment~\cite{albarelli_invasiveness_2023}. Intuitively, one may expect that improved precision comes at the cost of increased measurement backaction, especially if the number of impurities scales extensively with system size. This could be quantified using the recently developed thermodynamic description of decoherence~\cite{Popovic2021,popovic_thermodynamics_2023}. 

\begin{acknowledgments}
We are grateful to S.~Campbell, G.~Mihailescu, and A.~K.~Mitchell for insightful discussions, and we also thank S.~Campbell, R.~Onofrio, and A.~Purkayastha for useful comments on the manuscript. We acknowledge financial support from a Royal Society-Science Foundation Ireland University Research Fellowship (URF\textbackslash R1\textbackslash 221571).
\end{acknowledgments}

\appendix
\section{Solutions to the single particle Schrödinger equations}
\label{single_part_schrodinger}
In this section, we solve the single-particle Schrödinger equation for a particle in a box, with no, one, or two delta potentials present. We will solve the equations 
\begin{equation}
    \hat{h}_\sigma \psi_n^{(\sigma)} = E_n^{(\sigma)}\psi_n^{(\sigma)}.
\end{equation}
for each $\sigma$. 

For $\hat{h}_{\downarrow\downarrow}$, this is simply given by
\begin{equation}
\label{unperturbed_wf}
    \psi_n^{(\downarrow\downarrow)}\equiv \psi_n = \sqrt{\frac{2}{L}} \sin\left(k_n(x+L/2)\right),
\end{equation}
and $E_n = \hbar^2k_n^2/2m$, with $k_n = n\pi/L$. 

We next consider the case of $\sigma = \downarrow\uparrow$. The eigenfunctions in this case are given by 
\begin{equation}
\label{eigenfunc_downup}
    \psi_n^{(\downarrow\uparrow)} = 
        \begin{cases}
            A_n \sin{\left(k_n^{\prime}(x+L/2)\right)},\;-L/2<x< x_0\\
            B_n \sin{\left(k_n^{\prime}(x-L/2)\right)},\quad L/2>x> x_0
        \end{cases}
\end{equation}
with energy
\begin{equation}
    E_n^{(\downarrow\uparrow)} \equiv E_n^\prime = \frac{\hbar^2k_n^{\prime 2}}{2m}
\end{equation}
and $k_n^\prime$ is given by the quantization condition
\begin{equation}
    \cot\left(k_n^\prime(x_0+L/2\right) -\cot\left(k_n^\prime(x_0 - L/2)\right) = \frac{1}{k_n^\prime a}.
\end{equation}
The coefficients $A_n$ and $B_n$ can be found by the normalization requirement. 

The solution for $\sigma = \uparrow\downarrow$ is similar to the one above. We can note immediately that the eigenfunctions should be mirrored versions of Eq. (\ref{eigenfunc_downup}) around the center of the box. Indeed we find that 
\begin{equation}
    \psi_n^{(\uparrow\downarrow)} = 
        \begin{cases}
            (-1)^{-n}B_n \sin{\left(k_n^{\prime}(x+L/2)\right)},\;-L/2<x< -x_0\\
            (-1)^{-n}A_n \sin{\left(k_n^{\prime}(x-L/2)\right)},\quad L/2>x> -x_0
        \end{cases}
\end{equation}
with the same eigenenergies as above. 

For $\sigma=\uparrow\uparrow$ we treat the even and odd eigenfunctions separately. For $n$ even we find that the solution can be written
\begin{equation}
    \psi_n^{(\uparrow\uparrow)} = 
        \begin{cases}
            C_n \sin{\left(k_n^{\prime\prime}x - \delta_n \right)},&-L/2<x<- x_0\\
            D_n \sin{\left(k_n^{\prime\prime}x\right)},& -x_0<x< x_0\\
            C_n \sin{\left(k_n^{\prime\prime}x + \delta_n \right)},& L/2>x> x_0
        \end{cases}
\end{equation}
where $k_n^{\prime\prime}= k_n - 2\delta_n/L$ and the quantization condition takes the form
\begin{equation}
    \cot(k_n^{\prime\prime}x_0) - \cot(k_n^{\prime\prime}x_0+\delta_n) = \frac{2}{k_n^{\prime\prime}a}.
\end{equation}
For $n$ odd, the eigenfunctions are given by 
\begin{equation}
    \psi_n^{(\uparrow\uparrow)} = 
        \begin{cases}
            C_n \cos{\left(k_n^{\prime\prime}x - \delta_n \right)},&\quad-L/2<x<- x_0\\
            D_n \cos{\left(k_n^{\prime\prime}x\right)},&\quad -x_0<x< x_0\\
            C_n \cos{\left(k_n^{\prime\prime}x + \delta_n \right)},&\quad L/2>x> x_0.
        \end{cases}
\end{equation}
with $k_n^{\prime\prime}$ the same as above. The quantization condition this time takes the form
\begin{equation}
     \tan(k_n^{\prime\prime}x_0+\delta_n) - \tan(k_n^{\prime\prime}x_0) = \frac{2}{k_n^{\prime\prime}a}.
\end{equation}
The eigenenergy is in both cases given by 
\begin{equation}
    E_n^{(\uparrow\uparrow)}\equiv E_n^{\prime\prime} = \frac{\hbar^2k_n^{\prime\prime 2}}{2m},
\end{equation}
and the coefficients $C_n$ and $D_n$ can be found by requiring normalized states. It is worth noting that one may run into numerical instabilities when solving for $k_n^{\prime\prime}$ and $\delta_n$. We implement some safety procedures to check the validity of our solutions. The phase shift should never be higher than $\pi$, and the magnitude of $\delta_n$ can oscillate but should be contained within a decaying envelope function going to zero as $n$ increases. If these conditions are not satisfied for any given pair $\delta_n$, $k_n^{\prime\prime}$, we find that rewriting the quantization conditions may help in solving the equations.

\section{Computational details for the decoherence functions}

\label{app:comp_details}
In this section, we give some computational details for evaluating eq. (\ref{levitov}). 

For each temperature $T$, we need to determine the chemical potential $\mu$. We do this by solving the equation 
\begin{equation}
    \tr \hat{n} = N_s,
\end{equation}
with $N_s$ the number of particles in our gas. The trace is computed over a large basis set. In our calculations, we used $\sim 10^4$ basis states.

We go to the thermodynamic limit by increasing the number of particles in the gas $N_s$ while keeping the density $\bar n = N_s/L$ fixed until convergence is reached. For the timescales we are considering, we find that using $N_s$ in the range of $500-1000$ gives good convergence avoiding finite size effects in the time scale we are interested in. 

We need to calculate the matrix elements of operators like 
\begin{equation}
    \hat{A} = 1-\hat{n} +\hat{n}e^{i\hat{h}_{\uparrow\downarrow}t/\hbar}e^{-i\hat{h}_{\downarrow\uparrow}t/\hbar}.
\end{equation}
This matrix is in principle infinite-dimensional, but we can get a good approximation by using a finite basis set of size $N$. We determine $N$ by requiring   
\begin{equation}
    \big|\sum_{i=1}^N f(E_i) -N_s\big| < \epsilon,
\end{equation}
with $f(E)$ is the Fermi-Dirac distribution of the unperturbed gas. We find that using $\epsilon = 10^{-3}$ gives good convergence. In Appendix \ref{single_part_schrodinger} we calculated the eigenfunctions and energies of the single particle operators. We insert the resolution of identity and get the matrix elements of the $N\cross N$ matrix
\begin{widetext}
\begin{equation}
        A_{nm} = (1-f(E_n))\delta_{nm} + 
        f(E_n)\sum_{i=1}^{N^\prime}\sum_{j=1}^{N^\prime}\sum_{k=1}^{N^{\prime\prime}}e^{i(E_i^\prime-E_j^\prime)t/\hbar}\braket*{\psi_n}{\psi_i^{(\uparrow\downarrow)}}\braket*{\psi_i^{(\uparrow\downarrow)}}{\psi_k}
         \braket*{\psi_k}{\psi_j^{(\downarrow\uparrow)}}\braket*{\psi_j^{(\downarrow\uparrow)}}{\psi_m}.
\end{equation}
\end{widetext}
We have to fix the size of the perturbed basis set $N^\prime$. We do this by the requirement of unitarity
\begin{equation}
    \sum_{n=1}^{N^\prime}\big|\braket*{\psi_m}{\psi_n^{(\uparrow\downarrow)}}\big|^2 >1-\epsilon,
\end{equation}
for any $m\leq N$. We then use this to determine the size of the unperturbed basis set we also insert $N^{\prime\prime}$ by requiring 
\begin{equation}
    \sum_{n=1}^{N^{\prime\prime}}\big|\braket*{\psi_m^{(\uparrow\downarrow)}}{\psi_n}\big|^2 >1-\epsilon,
\end{equation}
for any $m\leq N^\prime$. We find that using $\epsilon \sim 10^{-4}$ yields excellent convergence. In a completely analogous way, we can calculate the other 3 decoherence functions.

\section{Cumulant expansion for weak coupling and low temperature}
\label{app_cumulant}
In this section, we derive the behavior of the decoherence functions, both in the power-law and thermal regimes. We explicitly go through the derivation for $\nuuddu(t)$ and state the results for the other three. They can be found in a completely analogous way. Our results are valid in the weak coupling $k_Fa\ll 1$ and low temperature $T/T_F\ll 1$ regime. This calculation follows closely to the one discussed in the supplemental material of \cite{mitchison_situ_2020}.

\begin{figure}
    \centering
    \includegraphics[width=0.49\textwidth]{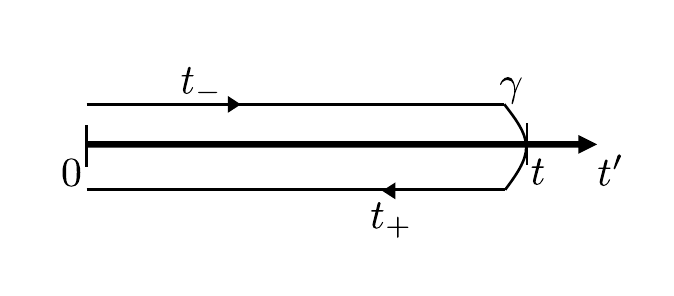}
    \caption{The contour $\gamma$ on which the time-ordering needs to be performed. In the time-ordering, operators on the lower branch are later than operators on the upper branch.}
    \label{contour_time_order}
\end{figure}

The starting point of our derivation is the many-body expectation value of eq. (\ref{decoherence_functions}) of the main text which for the decoherence function we consider reads
\begin{equation}
    \nuuddu(t) = \Big\langle e^{i\hat{H}_{\uparrow\downarrow} t/\hbar}e^{-i\hat{H}_{\downarrow\uparrow}t/\hbar}\Big\rangle.
\end{equation}
We might write the Hamiltonians appearing as $\hat{H}_\sigma = \hat H_0 + \hat V_\sigma$ with 
\begin{equation}
    \hat H_0 = \sum_k E_k c_k^\dagger c_k
\end{equation}
\begin{equation}
    \hat H_\sigma = \sum_{n,m}V_{nm}^{(\sigma)} c_n^\dagger c_m,
\end{equation}
where $c_n^\dagger$ creates a fermion in the state $\psi_n$ given in eq. (\ref{unperturbed_wf}). The interaction matrix $V_{nm}^{(\sigma)}$ has elements given by
\begin{equation}
\begin{split}
    V_{nm}^{(\sigma)} &= \int_{-L/2}^{L/2}\dd x\psi_n(x)\lambda\delta(x\pm x_0)\psi_m(x)\\ 
    &= \frac{2\lambda}{L}\sin\left(k_n x_0\pm\frac{\pi}{2} \right)\sin\left(k_m x_0\pm\frac{\pi}{2} \right),
\end{split}
\end{equation}
where $\lambda = -\hbar^2/ma$,  $k_n = n\pi/L$, and the plus (minus) sign is valid for $\sigma=\uparrow\downarrow$ ($\downarrow\uparrow$). 

To simplify notation we will from now denote $\nuuddu = |\nu| e^{i\phi}$. We can write this as a time-ordered exponential, which can be expanded in terms of time-ordered cumulants \cite{kubo_generalized_1962}
\begin{equation}
\label{cumulant_expansion}
\begin{split}
    |\nu| e^{i\phi} &=  \expval{\mathcal{T}\exp\bigg[\int_\gamma\dd t^\prime \frac{\hat{V}(t^\prime)}{i\hbar}  \bigg]}\\
    &\approx \exp\left[ \bigg\langle\int_\gamma \dd t^\prime \frac{\hat{V}(t^\prime)}{i\hbar}\bigg\rangle_c + \frac{\mathcal{T}}{2}\bigg\langle\bigg(\int_\gamma \dd t^\prime \frac{\hat{V}(t^\prime)}{i\hbar}\bigg)^2\bigg\rangle_c \right],
\end{split}
\end{equation}
where the symbol $\mathcal{T}$ denotes time ordering on the curve $\gamma$ shown in Fig. \ref{contour_time_order} such that operators on the branch $t_+$ occurs at a later time than operators on the $t_-$ branch. The integration is performed over this same curve. $\expval{\bullet}$ denotes the thermal expectation value with respect to the initial thermal state, and $\expval{\bullet}_c$ is the corresponding cumulant. The operator $\hat{V}(t)$ in the equation above takes the form
\begin{equation}
    \hat{V}(t) = 
    \begin{cases}
            \hat{V}_{\uparrow\downarrow}(t),\quad t\in t_-\\
            \hat{V}_{\downarrow\uparrow}(t),\quad t\in t_+
        \end{cases}
\end{equation}
 on the two branches of the curve $\gamma$. The operators are written in the interaction picture such that $\hat{V}_\sigma(t) = e^{i\hat{H}_0t/\hbar}\hat{V}_\sigma e^{-i\hat{H}_0t/\hbar}$. In the second line of Eq. (\ref{cumulant_expansion}) we have neglected terms of order $\mathcal{O}(\hat{V}^3)$.

By using the thermal expectation value $\expval{c_n^\dagger c_m} = f(E_n)\delta_{nm}$, with $f(E)$ being the Fermi-Dirac distribution, we find that the first cumulant is
\begin{equation}
\begin{split}
    &\int_\gamma \dd t^\prime \langle{\hat{V}(t^\prime)}\rangle_c = \int_0^t \dd t^\prime \frac{\langle{\hat{V}_{\uparrow\downarrow}(t^\prime)}\rangle_c}{i\hbar} + \int_t^0 \dd t^\prime \frac{\langle{\hat{V}_{\downarrow\uparrow}(t^\prime)}\rangle_c}{i\hbar}\\
    &=\frac{t}{i\hbar}\sum_n f(E_n)\left(V_{nn}^{(\uparrow\downarrow)} - V_{nn}^{(\downarrow\uparrow)} \right) = 0.
\end{split}
\end{equation}

To capture the decay of the decoherence function we calculate the second cumulant. We use the standard relation for Gaussian fermionic states $\langle{c_n^\dagger c_m c_k^\dagger c_l}\rangle_c = f(E_n)[1-f(E_m)]\delta_{nl}\delta_{mk}$ to get 
\begin{equation}
\begin{split}
    &\expval{\hat{V}_\sigma(t)\hat{V}_{\sigma^\prime}(t^{\prime})}_c\\
    &= \sum_{n,m} V_{nm}^{(\sigma)}V_{mn}^{(\sigma^\prime)}e^{i(E_n-E_m)(t-t^\prime)/\hbar}f(E_n)[1-f(E_m)].
\end{split}
\end{equation}
The terms with $n=m$ will contribute a term proportional to $t^2$, and will go to zero as $L^{-1}$, and can be neglected in the thermodynamic limit. We then use this result to compute the second cumulant. After integrating over the contour $\gamma$, being careful of the time ordering, we find that 
%\begin{widetext}
\begin{equation}
\label{ugly_decoherence}
 \begin{split}   &-\Gamma(t)\equiv-\frac{\mathcal{T}}{2}\bigg\langle\bigg(\int_\gamma \dd t^\prime \frac{\hat{V}(t^\prime)}{i\hbar}\bigg)^2\bigg\rangle_c\\ 
 &= \sum_{n\neq m}f(E_n)[1-f(E_m)]\frac{1-\cos[(E_n-E_m)t/\hbar]}{(E_n-E_m)^2}V_{nm}^2,
\end{split}
\end{equation}
%\end{widetext}
where we have defined $V_{nm} = V_{nm}^{(\uparrow\downarrow)}-V_{nm}^{(\downarrow\uparrow)}$. Note that Eq. (\ref{ugly_decoherence}) is a real number, so the decoherence function is real at least to second order in $\hat{V}$. This is unique for the $\nuuddu$, and will not hold true for the other decoherence functions.

To simplify Eq. (\ref{ugly_decoherence}), we introduce the spectral density 
\begin{equation}
\begin{split}
    J(\omega) &= \frac{1}{\hbar}\sum_{n,m}V_{nm}^2f(E_n)[1-f(E_m)]\delta(\hbar\omega +E_n-E_m)\\
    &=\frac{1}{\hbar}\int_0^\infty\dd E f(E)[1-f(E+\hbar\omega)]\\
    &\cross\sum_{n\neq m}V_{nm}^2\delta(E-E_n)\delta(\hbar\omega +E-E_m)
\end{split}
\end{equation}
representing the coupling strength to particle-hole excitations weighted by the finite temperature density of states. The decoherence rate then takes the form
\begin{equation}
\label{decoherence_integral}
    \Gamma (t) = -\fint_{-\infty}^\infty \dd \omega \frac{1-\cos(\omega t)}{\omega^2}J(\omega),
\end{equation}
where $\fint$ denotes the principal value integral excluding $\omega=0$. To make further progress we take the continuum limit, introducing the s-wave density of states 
\begin{equation}
    D_s(E) = \frac{1}{L}\sum_n\delta(E-E_n) = \frac{1}{\pi\hbar}\sqrt{\frac{m}{2E}}
\end{equation}
valid when $L\rightarrow\infty$. We replace $V_{nm}$ by $V(E)$ by letting 
\begin{equation}
    k_n\rightarrow k(E) = \sqrt{\frac{2mE}{\hbar^2}}.
\end{equation}
Doing this simplifies the spectral density to take the form
\begin{equation}
\label{spectral_density}
\begin{split}
    J(\omega)= \frac{2\lambda^2}{\hbar}\int_0^\infty &\dd E f(E)[1-f(E+\hbar\omega)]\\
    &\cross D_s(E)D_s(E+\hbar\omega)g(E,\omega).
\end{split}
\end{equation}
where we have introduced the function $g$ capturing the effect of the impurities on the gas. We can write it as
\begin{equation}
    g(E,\omega) = 1-\cos(2k(E)x_0)\cos(2k(E+\hbar\omega)x_0).
\end{equation}
We are interested in low temperatures such that $T\ll T_F$ and $\mu\approx E_F$. For $\hbar\omega\leq -E_F$, $J(\omega)$ is exponentially suppressed. If we for a second ignore $g(E,\omega)$ in Eq. (\ref{spectral_density}), we see that $J(\omega)\sim \sqrt{\omega\tau_F}$ for $\hbar\omega\geq E_F$. Going back to Eq. (\ref{decoherence_integral}), such contributions can be neglected in $\Gamma(t)$. Including $g(E,\omega)$, will only make $J(\omega)$ grow slower as a function of $\gamma$, and thus we can restrict ourselves to low frequencies $\hbar|\omega|\ll E_F$. In this regime, the function $f(E)[1-f(E+\hbar\omega)$ is sharply peaked around $E=E_F$, and we may replace $\sqrt{E(E+\hbar\omega}\rightarrow \sqrt{E_F(E_F+\hbar\omega}\approx E_F$. In $g(E,\omega)$ we make the replacement $E\rightarrow E_F$ and make a series expansion in $\hbar\omega/E_F$ in $k(E+\hbar\omega)$. We also introduce the interaction time $\tau_i = 2x_0/v_F$ as discussed in the main text to get the expression
\begin{equation}
    g(\omega) = 1-\cos(2\tau_i/\tau_F)\cos(2\tau_i/\tau_F + \tau_i\omega).
\end{equation}
The remaining integral in Eq. (\ref{spectral_density}) can then be computed to yield
\begin{equation}
    J(\omega) = \frac{1}{2}\alpha \omega g(\omega)\left[1+\coth(\hbar\beta\omega/2) \right],
\end{equation}
which is valid for $|\omega|<\Lambda$, where $\Lambda\sim E_F/\hbar$ is a cutoff-frequency. Here, we have introduced the dimensionless coupling strength $\alpha = (\pi k_Fa)^{-2}$.

Going back to Eq. (\ref{decoherence_integral}), we have to calculate the following integral
\begin{equation}
\begin{split}
\label{decoherence_two_integrals}
    \Gamma(t) = &-\frac{\alpha}{2}\fint_{-\Lambda}^\Lambda\dd \omega \frac{1-\cos\omega t}{\omega}g(\omega)[1+\coth(\hbar\beta\omega/2)]\\
    = &-\frac{\alpha}{2}\fint_{-\infty}^\infty\dd \omega \frac{1-\cos\omega t}{\omega}g(\omega)[1+\coth(\hbar\beta\omega/2)]\\
    &+\alpha\int_\Lambda^\infty\dd\omega \frac{1-\cos\omega t}{\omega}g(\omega),
\end{split}
\end{equation}
where in the third line we have used that $\coth(\hbar\beta\omega/2)\approx\text{sign}(\omega)$ for $|\omega|>\Lambda$, which is valid as long as $\hbar\beta\Lambda\gg 1$. The two integrals in the second equality of Eq. (\ref{decoherence_two_integrals}) can be evaluated by writing the cosines as complex exponentials and we end up with 9 terms of the form
\begin{equation}
\label{form_of_integrals}
    -\frac{\alpha}{2}\fint_{\infty}^\infty \dd \omega e^{iy\omega}[1+\coth(\hbar\beta\omega/2)]/\omega +\alpha \int_\Lambda^\infty\dd \omega e^{iy\omega}/\omega.
\end{equation}
For the first of these integrals, we extend $\omega$ to the complex plane, and depending on the sign of $y$ we close the contour in either the upper or lower plane, being careful going around the origin. We then employ the residue theorem, and end up with a geometric series over the residues at the poles at $\omega = i\omega_n$ with $\omega_n = 2n\pi/\hbar\beta$ the bosonic Matsubara frequencies. $n$ runs over $n=\pm1,\pm2,\dots$ where the + (-) sign is valid if $y>0$ ($y<0$) such that the integrand vanish on the edge of the contour. 

The second integral in Eq. (\ref{form_of_integrals}) can be explicitly evaluated by adding a small imaginary part to $y\rightarrow y+i\epsilon$ and taking the limit $\epsilon\rightarrow 0$. The integral will be proportional to the incomplete gamma function $\Gamma(0,iy)$. Combining this we obtain the following expression for the decoherence rate
\begin{equation}
\begin{split}
    &\Gamma(t) = -2\alpha \left[\ln{\frac{\hbar\Lambda\beta}{\pi}\sinh\left(\frac{\pi t}{\hbar \beta} \right)} -\text{Ci}(\Lambda t) + \gamma_E\right]\\
    &-\alpha\cos^2\frac{2\tau_i}{\tau_F}\ln{\frac{\left(1-e^{-\frac{2\pi\tau_i}{\hbar\beta}}\right)^2}{\left(1-e^{-\frac{2\pi(t+\tau_i)}{\hbar\beta}}\right)\left(1-e^{-\frac{2\pi|t-\tau_i|}{\hbar\beta}}\right)}} \\
    &+\alpha\cos^2\frac{2\tau_i}{\tau_F} \big[2\text{Ci}(\tau_i\Lambda) - \text{Ci}((t+\tau_i)\Lambda) -\text{Ci}(|t-\tau_i|\Lambda) \big]\\
    &+\alpha\cos\frac{2\tau_i}{\tau_F} \sin\frac{2\tau_i}{\tau_F} \big[ 2\text{Si}(\tau_i\Lambda) - \text{Si}((t+\tau_i)\Lambda)\\
    &\qquad\qquad\qquad\qquad -\text{Si}(|t-\tau_i|\Lambda)\big],
\end{split}
\end{equation}
where Ci and Si denote the cosine and sine integrals, $\gamma_E$ is the Euler constant and we have used the relation
\begin{equation}
\begin{split}
    &e^{i\theta}\Gamma(0,ix) + e^{-i\theta}\Gamma(0,-ix)\\ 
    &= -2\cos\theta \text{Ci}(x)-2\sin\theta(\text{Si}(x)-\pi/2).
\end{split}
\end{equation}
We are interested in how the decoherence behaves in time, so all constant terms will be neglected. The terms proportional to $\text{Ci}(\Lambda t)$, $\text{Ci}(\Lambda (t+\tau_i))$ and $\text{Si}(\Lambda (t+\tau_i))$ are oscillating, but will quickly become negligible compared to the other terms. The terms proportional to $\text{Ci}(|t-\tau_i|\Lambda)$ and $\text{Si}(|t-\tau_i|\Lambda)$ regulate the solution at $t=\tau_i$. If these terms did not appear, the decoherence function would go to zero at this point. We are left with a decoherence function of the following form
\begin{equation}
\begin{split}
    &\nuuddu(t)\propto \left[\frac{\hbar\beta}{\pi\tau_F}\sinh \frac{\pi t}{\hbar\beta} \right]^{-2\alpha}\\
    &\cross\left[\frac{\left(1-e^{-\frac{2\pi\tau_i}{\hbar\beta}}\right)^2}{\left(1-e^{-\frac{2\pi(t+\tau_i)}{\hbar\beta}}\right)\left(1-e^{-\frac{2\pi|t-\tau_i|}{\hbar\beta}}\right)}\right]^{-\alpha\cos^2\frac{2\tau_i}{\tau_F}}.
\end{split}
\end{equation}
Here, we got rid of the cut-off dependence by using that $(\Lambda/E_F)^{-2\alpha}\sim 1$. 

We find three regimes of the decoherence function. Firstly, if $t\ll\tau_i\ll\hbar\beta$, we have algebraic decoherence with an exponent $2\alpha$. This is the same as two independent impurities decohering. If we have $\tau_i\ll t\ll \hbar\beta$ the exponent changes to $-2\alpha(1-\cos(4 k_Fx_0)$. Thus the bath-induced interaction leads to suppressed decoherence. This effect is known as sub-decoherence \cite{palma_quantum_1996,reina_decoherence_2002}, and will be discussed more in the next paragraph. As the impurities are only able to produce low-energy, long-wavelength excitations, the decoherence slows down. Finally, for $t\gg \hbar\beta$ the decoherence is exponential $\nuuddu(t)\propto e^{-\gamma t}$ with $\gamma = \frac{2\alpha\pi}{\hbar\beta}$.

To see why the sub-decoherence occurs for $\nuuddu$ it is useful to rewrite the interaction Hamiltonian in the Bell-basis. It can easily be checked that the interaction Hamiltonian can be written as
\begin{equation}
\begin{split}
\label{bell_hamiltonian}
    \frac{\hat{H}_I}{\lambda} &= \frac{1}{2}\hat{I}\otimes(\hat{n}_1+\hat{n}_2)\\ 
    &+ \frac{1}{2}\left(\dyad{\Phi^+}{\Phi^-} + h.c. \right)\otimes(\hat{n}_1+\hat{n}_2)\\
    &+ \frac{1}{2}\left(\dyad{\Psi^+}{\Psi^-} + h.c. \right)\otimes(\hat{n}_1-\hat{n}_2),
\end{split}
\end{equation}
where $\hat{n}_1$ and $\hat{n}_2$ are density operators for the gas at positions $\pm x_0$. It is clear that decoherence functions like $\nuuddu$ couple to density differences in the gas. At long times, only low-frequency excitations are relevant for the decoherence function: in particular, at time $t$ only excitations with frequencies $\omega \lesssim 1/t$ contribute significantly. Low frequency generally entails long wavelength, and density excitations with wavelengths much greater than the impurity separation are not able to create an appreciable density difference at the positions of the impurities. In particular, excitations close to the Fermi surface have a wavelength $\lambda  \sim  2\pi v_F/\omega$, and therefore the decoherence signal slows down significantly when $t\gtrsim \tau_i$ after which time all relevant wavelengths exceed $\Delta x$.

Following the approach outlined in this section, we can also compute the behavior of the other decoherence functions. Here, we just state the results. For the decoherence function $\nuuddd$, the behavior is that of a single qubit in a thermal bath. For short times $t\ll\hbar\beta$ we have algebraic decay with exponent $\alpha$, while for long times $t\gg\hbar \beta$ we have exponential decay $\sim e^{-\alpha\pi t/\hbar\beta}$. 

For the decoherence function $\nuuudd$, we have the same behavior as $\nuuddu$ for short times, that is algebraic decay with exponent $2\alpha$, but for the intermediate regime, $\tau_i\ll t\ll\hbar\beta$ the exponent changes to $-2\alpha(1+\cos(4 k_Fx_0)$. As can be seen from Eq. (\ref{bell_hamiltonian}), this decoherence function couples to low-frequency excitations, leading to the phenomena of super-decoherence.Finally, in the limit $t\gg \hbar\beta$ the decoherence is exponential $\sim e^{-2\alpha\pi t/\hbar\beta}$.

Finally, for $\nuuuud$, the low-$T$ weak coupling expansion breaks down. The derivation of the cumulant expansion of $\nuuuud$ will be almost identical to that given above, the only change being that $\hat{V}^{(\downarrow\uparrow)}\rightarrow \hat{V}^{(\uparrow\uparrow)}$, where
\begin{align}
    \hat{V}^{(\uparrow\uparrow)}_{mn} = \hat{V}^{(\uparrow\downarrow)}_{mn} + \hat{V}^{(\downarrow\uparrow)}_{mn}
\end{align}
are the matrix elements of the interaction Hamiltonian when both impurities are in the state $\ket{\uparrow}$. Looking at Eq. (\ref{ugly_decoherence}), this means that the $V_{nm}$ that appears will change into $V_{nm} = V_{nm}^{(\uparrow\downarrow)}-V_{nm}^{(\uparrow\uparrow)} = -V^{(\downarrow\uparrow)}_{mn}$. This in turn means that the effect of the first impurity interacting with the gas cancels out, and we end up with essentially the same behavior for $\nuuuud$ as for $\nuuddd$, i.e.~a single impurity interacting with a bath of fermions. In order to capture the slowing down of the decoherence observed in Fig. \ref{dynamics_of_probes} of the main text, we need to consider higher-order cumulants.

\section{Quantum correlations}
\label{app:quant_corr}
In this section, we investigate the nature of the correlations emerging between the two impurities due to the bath-mediated interactions between them. 

The first and most obvious check for quantum correlations is to look for entanglement between the impurities. It is well known that for a two-qubit system the Peres-Horodecki criterion \cite{peres1996separability,horodecki1997separability} is necessary and sufficient for a state to be separable. The criterion states that the partial transpose of the density matrix must be positive for the state to be separable. We find that our two-qubit state remains separable for all times and for all parameters. 

However, quantum correlations may exist beyond entanglement (see e.g. \cite{Adesso_2016} for a comprehensive review). It is well established in the literature that the total amount of correlations between subsystems 1 and 2 is given by the mutual information (Eq. (\ref{mutual_info}) in the main text). One measure of genuine quantum correlation is the quantum discord \cite{ollivier2001quantum, Henderson_2001}, defined as the difference between mutual information and classical correlations
\begin{equation}
    \mathcal{D}(1:2) = I(1:2) - \mathcal{C}(1:2).
\end{equation}
This measure is in general not symmetric under the permutation of the subsystems, however, in our case it is. We can also think of the quantum discord as the amount of mutual information of subsystems 1 and 2 that cannot be extracted by local measurements on one of the subsystems. Thus, it becomes clear that the classical correlations can be defined as the maximum change in entropy of subsystem 1 upon a local projective measurement $\{\hat{\Pi}^2_i\}$ on subsystem 2
\begin{equation}
    \mathcal{C}(1:2) = S(\hat\rho_1) - \min_{\{\hat{\Pi}^2_i\}}\sum_i p_iS(\hat\rho_{1|2=i}),
\end{equation}
with $\{p_i, \hat\rho_{1|2=i}\}$ being the post measurement ensemble 
\begin{equation}
    p_i = \text{tr}\left[ (I\otimes\hat{\Pi}_i^2)\hat\rho(I\otimes\hat{\Pi}_i^2)^\dagger\right],
\end{equation}
\begin{equation}
    \hat\rho_{1|2=i} = \frac{(I\otimes\hat{\Pi}_i^2)\hat\rho(I\otimes\hat{\Pi}_i^2)^\dagger}{p_i}.
\end{equation}
We can immediately see that the quantum discord is in general difficult to calculate due to the optimization required over the local projection operator. Luckily, for a two-qubit system, analytical results have been obtained. In order to compute the quantum discord we follow the procedure outlined in ref. \cite{girolami2011quantum}, where the optimization has been replaced with solving transcendental equations involving the elements of the density matrix of the system, i.e. the decoherence functions calculated in the main text. The equations can easily be implemented on a computer, but are quite involved. We refer the interested reader to \cite{girolami2011quantum} for further details on how to calculate the quantum discord.

\bibliographystyle{apsrev4-2}
\bibliography{mark_references.bib,references.bib}

\clearpage 

\end{document}